\documentclass[fleqn,usenatbib]{mnras}

\usepackage{newtxtext,newtxmath}

\usepackage[T1]{fontenc}
\usepackage{ae,aecompl}

\usepackage{graphicx}	

\usepackage{xspace}
\newcommand{\numax}{\ensuremath{\nu_{\textrm{max}}}\xspace}
\newcommand{\Kepler}{\ensuremath{\emph{Kepler}}\xspace}

\usepackage{comment}
\usepackage{subfig}
\usepackage{etoolbox}

\makeatletter
\patchcmd\@combinedblfloats{\box\@outputbox}{\unvbox\@outputbox}{}{%
    \errmessage{\noexpand\@combinedblfloats could not be patched}%
}%
\makeatother

\title[Clumpiness]{Clumpiness: Time-domain classification of red-giant evolutionary states}

\author[J. S. Kuszlewicz et al.]{
James S. Kuszlewicz,$^{1, 2}$\thanks{E-mail: kuszlewicz@mps.mpg.de}
Saskia Hekker,$^{1,2}$
Keaton J. Bell$^{3,4}$
\\
$^{1}$Max-Planck-Institut f\"{u}r Sonnensystemforschung, Justus-von-Liebig-Weg 3, 37077 G\"{o}ttingen, Germany 
\\
$^{2}$Stellar Astrophysics Centre, Department of Physics and Astronomy, Aarhus University, Ny Munkegade 120, DK-8000 Aarhus C, Denmark \\
$^{3}$DIRAC Institute, Department of Astronomy, University of Washington, Seattle, WA-98195, USA\\
$^{4}$NSF Astronomy and Astrophysics Postdoctoral Fellow and DIRAC Fellow\\
}
\date{Accepted XXX. Received YYY; in original form ZZZ}

\pubyear{2020}

\begin{document}
\label{firstpage}
\pagerange{\pageref{firstpage}--\pageref{lastpage}}
\maketitle

\begin{abstract}
Long, high-quality time-series data provided by previous space-missions such as CoRoT and \textit{Kepler} have made it possible to derive the evolutionary state of red-giant stars, i.e. whether the stars are hydrogen-shell burning around an inert helium core or helium-core burning, from their individual oscillation modes. We utilise data from the \textit{Kepler} mission to develop a tool to classify the evolutionary state for the large number of stars being observed in the current era of K2, TESS and for the future PLATO mission. These missions provide new challenges for evolutionary state classification given the large number of stars being observed and the shorter observing duration of the data. We propose a new method, \texttt{Clumpiness}, based upon a supervised classification scheme that uses ``summary statistics'' of the time series, combined with distance information from the Gaia mission to predict the evolutionary state. Applying this to red giants in the APOKASC catalogue, we obtain a classification accuracy of $\sim91\%$ for the full 4 years of \emph{Kepler} data, for those stars that are either only hydrogen-shell burning or also helium-core burning. We also applied the method to shorter \emph{Kepler} datasets, mimicking CoRoT, K2 and TESS achieving an accuracy $>91\%$ even for the 27 day time series. This work paves the way towards fast, reliable classification of vast amounts of relatively short-time-span data with a few, well-engineered features.
\end{abstract}



\section{Introduction}

Stars in the red clump (RC), i.e.\ low-mass core-helium-burning (CHeB) stars, are a prominent feature in the colour-magnitude diagram of, for instance, open or globular clusters due to their narrow luminosity distribution. This narrow distribution is a result of the near-constant core mass at helium ignition of stars with degenerate cores. Red-clump stars are widespread throughout the Galaxy and are used as distance indicators \citep[e.g.][]{1998ApJ...494L.219P,1998MNRAS.301..149G,2008A&A...488..935G}. Furthermore, their function as standard candles provides the possibility to use the parallaxes from the Gaia mission \citep{2016A&A...595A...1G,2018A&A...616A...1G} to calibrate the absolute magnitude of the red clump \citep[e.g.][]{2017MNRAS.471..722H,10.1093/mnras/stz1092,2019arXiv191000398C}. The red clump has also been used as a benchmark to investigate possible systematics in the parallax measurements when combined with, for example, asteroseismology \citep[e.g.][]{2017A&A...598L...4D, 2017ApJ...844..102H, 2019A&A...628A..35K, 2019ApJ...878..136Z}.

The identification of field red-clump stars is not trivial because they have very similar surface properties to red-giant-branch (RGB) stars. To add to this, not all field stars are at the same distance and so the typical ``clump'' seen in the colour-magnitude diagram of clusters is smeared out. There are also additional dependencies on parameters such as extinction and metallicity \citep[e.g.][]{2001MNRAS.323..109G}. \cite{2014ApJ...790..127B} identified RC stars in APOGEE data by not only searching luminosity space, but also colour-metallicity-surface gravity-effective temperature space in order to reduce the amount of contamination from red-giant-branch stars. 

With the advent of CoRoT \citep{2006ESASP1306...33B} and \emph{Kepler} \citep{2010Sci...327..977B} the evolutionary state of red-giant stars can now also be classified through asteroseismology \citep[e.g.][]{2011Natur.471..608B,2011A&A...532A..86M,2012A&A...541A..51K,2013ApJ...765L..41S,2014A&A...572L...5M,2016A&A...588A..87V,2017MNRAS.466.3344E,2017EPJWC.16004006H,2018MNRAS.476.3233H}. The majority of these methods rely on the detection of individual mixed oscillation modes (those that have a gravity-mode-like character in the stellar core and pressure-mode-like character in the convective envelope) in the power spectrum. These modes are indeed resolved for 4 years of \emph{Kepler} data (hereby defined as long datasets). However, for the shorter 80-day datasets of K2 \citep{2014Howell_k2} or the 27-day time-series of TESS \citep{2015JATIS...1a4003R} (hereby referred to as short datasets) the frequency resolution is coarser, hampering the evolutionary state classification via mixed modes. To distinguish between RC and RGB stars it is therefore important to have classification methods that do not rely on the determination of individual mixed oscillation modes \emph{a priori}. This will also be of importance for future space missions such as PLATO \citep{2014ExA....38..249R}.

Evolutionary state determination does not have to be undertaken in the frequency domain and we show in this work how the time-series data can be used directly. In this way, we are not limited by the reduced frequency resolution in the power spectra for shorter time-series data. 

It is common practice to classify the variability of stellar lightcurves in the time domain using machine-learning techniques. Unsupervised techniques (those where the stellar classification is not known beforehand) have been applied to intrinsically variable stars (e.g Cepheids/RR Lyraes, delta-Scuti pulsators etc.) and extrinsically variable stars (e.g. eclipsing binaries) in K2 data by \cite{2015A&A...579A..19A, 2016MNRAS.456.2260A} who used a combination of self-organising maps (to extract salient features from the data) and random forests to classify the data. \cite{2018MNRAS.474.3259V} used a different approach to classify variable stars, applied to data from OGLE \citep{1996AcA....46...51U, 2008AcA....58..329U}, MACHO \citep{1997ApJ...486..697A, 2003ApJ...598..597A} and \emph{Kepler}, based upon the similarity between different time-series using a variability tree, which ranks lightcurves by their similarity. \cite{2018NatAs...2..151N} took a different approach still and successfully used a ``novel'' recurrent autoencoder structure to classify variable stars with irregularly-sampled data from ASAS \citep{2002AcA....52..397P}. The advantage of such a method is that no features are chosen a priori and the autoencoder instead extracted them automatically. This method was only optimised for periodic variables. \cite{2015MNRAS.451.3385K} chose yet another method and adopted featureless classification by decomposing data into a mixture of gaussians and using a distance matrix to classify similar targets, with applications to OGLE and ASAS. Many of the cases mentioned above used short or sparsely sampled datasets, whereas in the case of \emph{Kepler} data the sheer amount of data collected can be a bottleneck in the extraction of important features. For many types of intrinsically variable stars the periods of oscillation dictate that densely sample time-series are necessary to resolve the underlying signal. 

In this work we introduce \texttt{Clumpiness}, an approach aimed at deducing the evolutionary state of red giants in the time domain, whilst retaining interpretability and computational efficiency to create a classification tool that performs as well as possible on both long and short time-series data using the fewest number of features. 

\section{Features}\label{sec:features}

A key component of any machine learning scheme is the set of features that are extracted from the data. The features form the backbone of the analysis and are what the chosen algorithm will use to predict the class an object belongs to. In our case this is the evolutionary state of a red-giant star.

There are a number of packages in the literature for computing very general features for time-series classification, for example, \texttt{tsfresh} \citep{2016arXiv161007717C}, FATS (Feature Analysis for Time Series; \citealt{2015arXiv150600010N}); UPSILoN \citep{2015ascl.soft12019K}  and PTSA\footnote{https://github.com/compmem/ptsa} to name but a few. Rather than computing a large number of very general features, we instead compute a small number of features in this work that contain relevant information about the evolutionary state of the stars in question.  

\subsection{Time-series features}

In this work we focussed on red giants that show solar-like oscillations which are excited and damped by near-surface turbulent convection. The near-surface turbulent convection displays itself as granulation which is present in the power spectrum as a frequency-dependent signal (commonly modelled as red noise). Whereas the solar-like oscillations are only present in the power spectrum at specific frequencies, as defined by the underlying stellar properties. The dominant process contributing approximately 85\% of the variability power in the time-series is the granulation, determined by integrating separately the granulation and oscillations components of models fit to the power spectrum. However, both the granulation and oscillation signals can provide insights into the evolutionary state of the star.

\subsubsection{Time-series MAD}

The variance of time-series data of a star with a convective envelope containing granulation and solar-like oscillations scales with \numax \citep[the frequency of maximum oscillation power,][]{2012A&A...544A..90H}, which in turn scales with the stellar surface gravity \citep[e.g.][]{1991ApJ...368..599B,1995A&A...293...87K}.

To capture a measure of the variance we compute the MAD (median absolute deviation from the median)
\begin{equation}
	\textrm{MAD}(X_{0}) = \textrm{median}\left(\left|X_{0,i}-\textrm{median}(X_{0})\right|\right),
\label{eqn:MAD}
\end{equation}
where $X_{0}$ represents the whole time-series, and $X_{0,i}$ is a single data point from that time-series. We prefer the MAD over the variance as a feature because it is more resistant to outliers \citep[e.g.][]{LEYS2013764}\footnote{Note that the MAD can be converted to the variance by assuming that the data are Gaussian-distributed, which may not always the case.}.

\subsubsection{MAD of time-series first differences}

Alongside computing the MAD of the time-series, we also look at the MAD of the first differences, where we define the first differences $X_{1}$ of a time-series $X_{0}$ to be

\begin{equation}
    X_{1,i} = X_{0,i} - X_{0,i-1}.
\label{eqn:first_diffs}
\end{equation} 

The first differences remove any long term trends from the data revealing additional timescales in the data. The MAD (Eq.~\ref{eqn:MAD}) of the first differences provides information regarding the magnitude of the rate of change of the signal. Therefore stars with longer period granulation will have lower first-difference MAD values compared to those with shorter period granulation. The MAD of the first differences acts as a rudimentary high-pass filter and removes power from the low-frequency regime of the signal. This will, in general, suppress the contribution of the granulation to the signal and enhance the contribution from the oscillations (and possibly the white noise if the level is very high). This offers a different view of the data that is complementary with taking the MAD of the time-series.

\subsubsection{Normalised number of zero-crossings}\label{sec:zc}
To capture the dominant timescale we compute the normalised (by the number of data points) number of zero-crossings in the time-series data. This normalisation assumes that the time stamps where there are no data points (i.e. during gaps) are dropped and so the number of data points used to normalise the data are the actual number of data points irrespective of their distribution in time, i.e. in the case of gaps. This normalisation is to place all stars onto the same scale enabling direct comparisons between datasets of different length. We also correct the normalised number of zero-crossings for gaps in the data if needed (see Appendix~\ref{sec:fill} for more information).

We compute the zero-crossings by first constructing a ``clipped'' time-series, $Z_{0}$, from the original time-series $X_{0}$ of length $N$ such that
\begin{equation}
    Z_{0,i} = 
    \begin{cases}
        1 & \text{if } X_{0,i} \geq 0, \\
        0 & \text{if } X_{0,i} < 0. \\
    \end{cases}
\end{equation}
The number of zero-crossings, $D_{0}$, is then given by the number of value changes in the new ``clipped'' time-series, computed as
\begin{equation}
    D_{0} = \sum_{i=1}^{N-1}\left(Z_{0,i} - Z_{0,i-1}\right)^{2},
\label{eqn:D}
\end{equation}
where $0 \leq D_{0} \leq N-1$.

The total number of zero-crossings are normalised to obtain the relative number of zero-crossings for a time-series of a given length, as described in Appendix~\ref{sec:fill}.

\subsubsection{Signal coherency}\label{sec:psi}

We can expand on the idea of the number of zero-crossings in the time-series by going to higher-order differences, i.e. first differences of the time-series (Eq.\ref{eqn:first_diffs}), first differences of the first differences and so on.  We compute the ``clipped" time-series of the higher-order crossings in exactly the same way as for the original time-series data, as given by Eq.~\ref{eqn:D}, except now we use the time-series of higher-order differences as an input. The time-series of the $k$th higher-order crossings is defined by
\begin{equation}
    X_{k,i} = X_{k-1,i} - X_{k-1,i-1}.
\end{equation}
Denoting the original ``clipped'' time-series as $Z_{0}$ then we define the ``clipped'' time-series of the $k$th higher-order crossing time-series by $Z_{k}$ such that
\begin{equation}
    Z_{k, i} = 
    \begin{cases}
        1 & \text{if } X_{k,i} \geq 0, \\
        0 & \text{if } X_{k,i} < 0. \\
    \end{cases}
\end{equation}
The number of higher-order crossings becomes
\begin{equation}
    D_{k} = \sum^{N-1}_{i=1}(Z_{k,i} - Z_{k,i-1})^{2},
\label{eqn:hoc_zc}
\end{equation}

We combine the information contained within the number of higher-order crossings to maximise the information content of this set of features. Solar-like oscillations and granulation are stochastically-driven with lifetimes/timescales that change depending on the size of the star. For a fixed length of time-series data, the signal from a star that is more evolved (i.e. with longer granulation timescales) will appear more coherent than that of a star that is less evolved. In contrast to the more coherent signal as a star evolves, the incoherent nature of a white-noise signal forms the lower boundary of the coherency measure. Therefore we will make use of the higher-order crossings to compute a measure of the coherency of the time-series.

To compute a measure of the coherency we look at the increments of the higher-order crossings \citep{BAE199675}, i.e. the rate of change of the higher-order crossings as a function of their order. These are defined as
\begin{equation}
    \Delta_{k} = 
    \begin{cases}
        D_{0} & \text{if } k=0, \\
        D_{k} - D_{k-1} & \text{if } k=1,...,N-2, \\
        (N-1) - D_{k-2} & \text{if } k=N-1. 
    \end{cases}
\end{equation}
\cite{doi:10.1093/biomet/68.2.551} have shown that for $N\geq300$ (where $N$ is the number of data points in the time-series) the number of higher-order crossings (HOCs) increases for $k\leq7$. We can therefore assuming that our HOCs are ordered such that the increments by which they increase are always positive. \cite{10.2307/2240904} also showed that when $k=8$, $D_{k}/N$ (the high-order crossings normalised by the number of data points $N$) is already as large as 0.9. As a result the discriminatory power is greatly reduced when $k$ is larger because $D_{k}/N$ levels off and slowly approaches unity. In this work we found that $k=5$ is an acceptable upper limit that balances discriminatory power with computation time.

The increments are combined into the coherency measure $\psi^{2}$
\begin{equation}
    \psi^{2} = \sum_{k=0}^{K-1} \frac{\left(\Delta_{k}-\phi_{k}\right)^{2}}{\phi_{k}}.
\end{equation}
For $k=0,...,K$, where $K=5$ the increments simulated for a white noise time-series of length 100,000 points are given by $\phi_{k}=(0.167, 0.066, 0.038, 0.025, 0.018)$.

The coherency measure $\psi^{2}$ provides a way to discriminate between white-noise-like signals and those that are more coherent. For instance, for a purely sinusoidal (coherent) signal the increments $\Delta_{k}$ will be zero for all $k$. This is because the period of the signal does not change when the differences are taken and so 
\begin{equation}
    \psi^{2}_{\mathrm{sin}} = \sum_{k=0}^{K}\phi_{k}.
\end{equation}
For a completely incoherent signal (i.e. white noise) the increments are equal to $\phi_{k}$ because they are generated from the same underlying process and therefore
\begin{equation}
    \psi^{2}_{\mathrm{white}} = 0.
\end{equation}
Therefore, the upper and lower bounds of $\psi^{2}$ are defined such that $0\leq \psi^{2} \leq \sum_{k=0}^{K}\phi_{k}$. Note that we normalise $\psi^{2}$ in the same way as the number of zero-crossings (see Appendix~\ref{sec:fill}).

\subsection{Absolute K-band Magnitude}

Alongside information gathered from the time-series we include external information that contributes to improving the classification. We mentioned in the introduction that the red clump forms a well-defined region in luminosity and therefore absolute-magnitude space, so leveraging this information is particularly helpful. We therefore use parallax data from the Gaia mission \citep{2016A&A...595A...1G,2018A&A...616A...1G}. Rather than using the parallax as a feature we link this back to an intrinsic property of the star, namely the absolute magnitude, that can help distinguish RGB from RC. We choose to compute the absolute magnitude in the K-band because the contribution of interstellar extinction is greatly reduced and in most cases negligible. This reduces the effect of any possible sources of bias in the dust maps used calculate the absolute magnitude accurately.

The absolute magnitude is calculated in the K-band according to 
\begin{equation}
    M_{K_{\mathrm{s}}} = m_{K_{\mathrm{s}}} - \mu_{0} - A_{K_{\mathrm{s}}}.
\label{eqn:absmag}
\end{equation}
The apparent K-band magnitude of the star $m_{K_{\mathrm{s}}}$ is taken from the Two Micron All Sky Survey (2MASS; \citealt{2003yCat.2246....0C, 2006AJ....131.1163S}). We compute the interstellar extinction in the K-band, $A_{K_{\mathrm{s}}}$, from the 3D dust map of \cite{2015ApJ...810...25G}. The distance modulus, $\mu_{0} = 5\log_{10}(d) - 5$ where the distance, $d$, (in parsecs) is taken from \cite{2018AJ....156...58B}. We use a 4$''$ cross-match between the asteroseismic and Gaia datasets, as given by \url{gaia-kepler.fun}, and take the brightest G-band magnitude source in the case of multiple Gaia sources for a single asteroseismic object. We adopt the following selection criterion \citep{2018A&A...616A...2L} to ensure that the stars that satisfy this have well-derived parallaxes
\begin{equation}
\sqrt{\frac{\chi^{2}}{\nu}} < 1.2\times\max\left\{1, \exp\left[-0.2\left(G-19.5\right)\right]\right\},
\end{equation}
where $\chi^{2}$ is the goodness-of-fit of the single-star astrometric model as provided by Gaia, $\nu$ is the number of degrees of freedom in the astrometric model and $G$ is the mean Gaia magnitude. If a star does not satisfy the above criterion then it is not included in the training data (see next section) for the full classification algorithm, but is instead included in the training of a version of the algorithm without $M_{K_{\mathrm{s}}}$ included as a feature.

\begin{figure*}
\includegraphics[width=1.75\columnwidth]{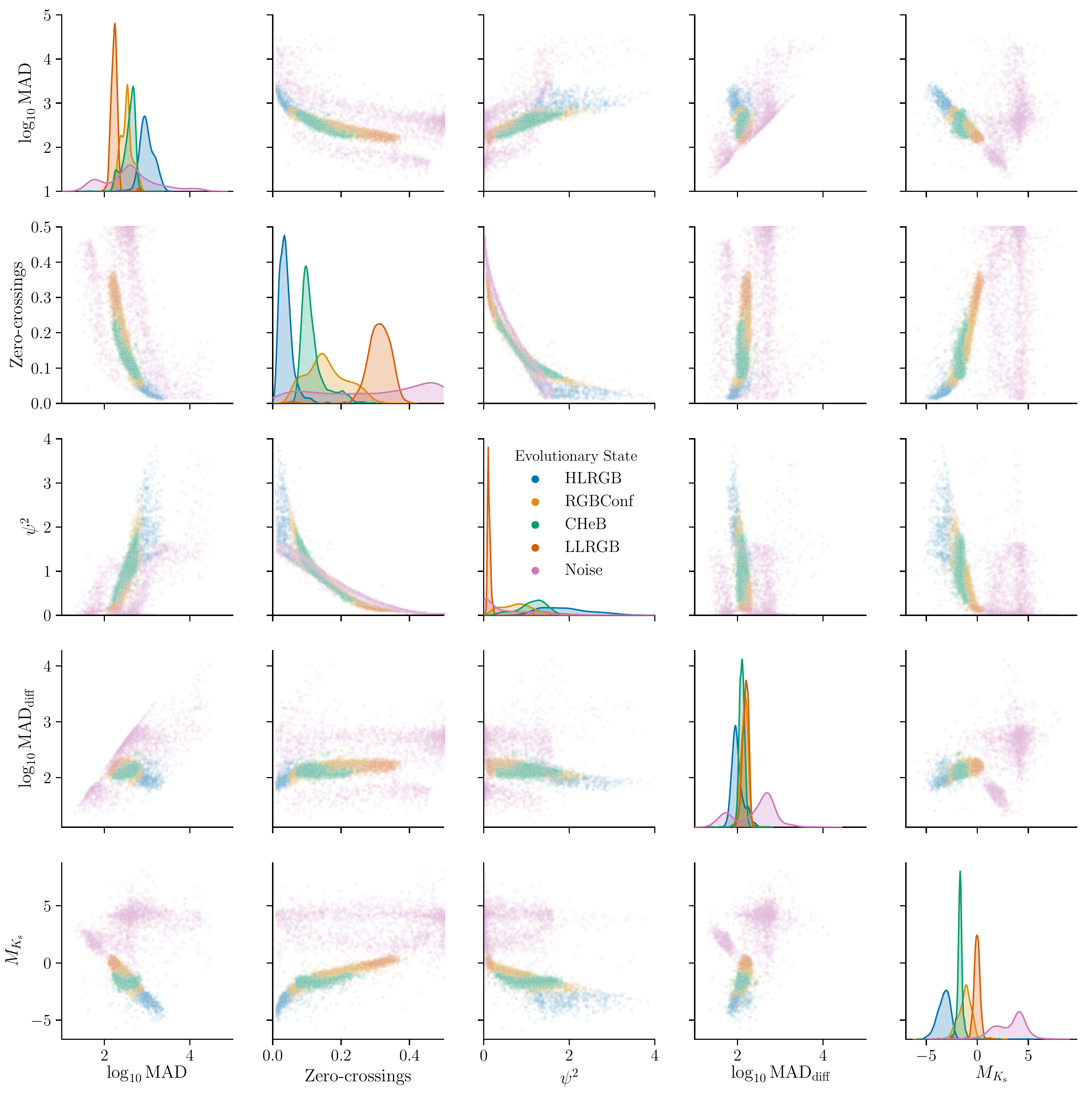}
\caption{A pairplot showing kernel density estimates (KDE) of all of the features (section \ref{sec:features}) broken down into individual classes as shown on the diagonal: high-luminosity red-giant branch (HLRGB), red-giant-branch stars that overlap with core-helium burning stars (RGBConf), core-helium burning (CHeB), low-luminosity red-giant-branch stars (LLRGB) and the noise class. The relationships between each set of features are shown on the off-diagonal panels.}
\label{fig:features}
\end{figure*}

\section{Training Data}

In this work we use red giants from the APOKASC sample \citep{2018ApJS..239...32P} observed by \emph{Kepler} with long-cadence (29.42 min) timeseries observations as training data. This is a sample of well studied red-giant stars that have consolidated evolutionary states from four different methods \citep{2019MNRAS.489.4641E}. We also include the sample of 472 main-sequence and sub-giant stars from \cite{2014ApJS..210....1C} and a random sample of 1500 \emph{Kepler} Objects of Interest (KOIs) from the NASA Exoplanet Archive \citep{ExArc2013} with $\log g > 3.5$ in the training data. These stars are not expected to show solar-like oscillations in the frequency range associated with long-cadence observations but are common contaminants in red-giant samples \citep[e.g.][]{2016MNRAS.463.1297Y} and act here as a representation of those contaminants that could find their way into such samples.

\subsection{Time-series preparation}

All the time-series used as training data are from long-cadence \emph{Kepler} observations produced using the pipeline developed by \cite{2010ApJ...713L.120J} with an additional Savitzky-Golay filter of width 20 days applied to remove any residual long-term trends. We use the \texttt{lightkurve} package \citep{2018ascl.soft12013L} for all preprocessing of the data. There are regular gaps in the time-series caused by the desaturation of the reaction wheels on the \emph{Kepler} spacecraft, typically the duration of one long-cadence (29.42 min) observation every three days. These were filled using linear interpolation according to \cite{2014A&A...568A..10G}. The data are then concatenated if segments of the data are separated by a long gap (over 20 days) following \cite{2010A&A...520A..60H}.

Alongside the full 4 years of data, we also produce reduced length datasets that correspond to the time-base of other missions, such as 80 days for K2 \citep{2014Howell_k2}, 180 days for CoRoT \citep{2006ESASP1306...33B} and 27 days for TESS \citep{2015JATIS...1a4003R}. For TESS we consider only the shortest length datasets as the limiting case in terms of dataset length.

To prepare shorter length datasets for algorithm training, we cut down the raw (now concatenated) data into chunks of the desired length (in time) and filter them with a Savitzky-Golay filter of width 10 days for the 180 day datasets, 5 days for 80 day datasets and 3 days for 27 day datasets. These filters mimic the detrending that would be used for the shorter datasets. In the case of the shortest length of 27 days, a linear trend is subtracted after the filtering to account for any possible trends in the data as a result of the short baseline. If any chunk of data has a duty cycle lower than 0.5 (i.e. more than 50\% of data are missing), then the chunk is discarded. This rarely happens since the data are concatenated in the presence of large gaps and so in most cases the duty cycle is above 0.9. Due to the stochastic nature of solar-like oscillations we are essentially gaining more independent training time-series when cutting the longer time-series down. This is due to the fact that the phase is incoherent in solar-like oscillators.

There is the possibility for a time-series of a given star to be contaminated by a nearby star due to the mask chosen for the extraction of the data. We do not want contaminated time-series to be used in our training set since this will confuse the classification algorithm. To remove contaminated time-series, we use a metric based on the quarter-by-quarter variance, $\sigma^{2}_{q,i}$. We use the following metric on the filtered data (see above)
\begin{equation}
	c = \mathrm{median}\left(\sum_{i=2}^{N_{\mathrm{quarters}}}\left|\sigma^{2}_{q,i}-\sigma^{2}_{q,i-1}\right|\right).
\end{equation}
This is the median of the absolute values of the first-differences of the quarter-by-quarter variances. When $\log_{10}c > 2.5$, which is determined empirically, the time-series is considered contaminated and is therefore no longer used.

\subsection{Class labels}

The evolutionary states for the red giants that make up our ground truth labels are taken from the consolidated classifications of the APOKASC sample by \cite{2019MNRAS.489.4641E}. Based on their classification and the inclusion of the main-sequence/KOI targets, we define three classes: red-giant branch (RGB), core-helium burning (CHeB) stars, and main-sequence stars/KOIs (Noise). 

In order to aid the classifier and given that the RGB class stretches over a wide range of parameter space, we split up the RGB class (for training purposes) into three sub-classes
\begin{itemize}
    \item Low-luminosity RGB (LLRGB) with  \numax$>130\mu$Hz.
    \item High-luminosity RGB (HLRGB) with  \numax$<15\mu$Hz.
    \item Confusion region RGB (ConfRGB) with $15<\;$\numax$<130\mu$Hz.
\end{itemize}
In this way the classifier can focus on disentangling CHeB stars from RGB stars in the region where they overlap (the so-called confusion region). We do not disentangle the low-mass red-clump stars from the higher-mass stars in the secondary clump. However, this could be performed afterwards using the probabilities from the classifier as an initial guide. Final probabilities are produced for the three main classes RGB, CHeB and noise.

\section{Supervised Classification}

The task of inferring the stellar evolutionary states from a set of derived features constitutes a supervised classification problem. In Fig.~\ref{fig:features} we plot each feature against every other for our training set, coloured by class label. This shows that the interactions between our features are not necessarily linear. They may be linear in log-space or under a more complicated transformation. In order to take advantage of this we use an ensemble algorithm \texttt{xgboost} \citep{2016arXiv160302754C}. An ensemble algorithm creates a sequence of weak classifiers whose individual performances are only slightly better than random guessing. When these weak classifiers are combined they produce a single strong classifier \citep[e.g.][]{Friedman2000,ESL}.

The \texttt{xgboost} algorithm is a variant of ensemble algorithms that falls under the umbrella of gradient-boosting methods. In boosting algorithms, models are constructed sequentially to correct the errors from the existing models until no further improvement can be made (according to some chosen criterion). In \texttt{xgboost} these models are decision trees. A gradient-boosting algorithm is a boosting algorithm where a gradient-descent algorithm is used to minimise the objective function (e.g. negative log-likelihood) to train the model. At each iteration, the data that are misclassified by the previous classifier in the sequence are upweighted such that the next classifier focusses more on those data. 

In order to fit the model to the data during training, an objective function needs to be chosen. This is the function that we are minimising via gradient descent in order to find the best fit to the data. We adopt the multi-class logarithmic loss since our problem is multi-class classification. This essentially acts as a measure of the degree of similarity between the ground truth and the predicted class probability which is defined by
\begin{equation}
    \ln\mathcal{L}=-\frac{1}{N}\sum_{i=1}^{N}\sum_{j=1}^{M}y_{ij}\ln p_{ij} + \sum_{k}\Omega(f_{k}).
\label{eqn:logloss}
\end{equation}
The double summation is over each of the $N$ time-series in our training set and $M$ classes, $y_{ij}$ is a binary indicator showing whether the label $j$ is the correct classification of the $i$th data point and $p_{ij}$ is the probability of the $i$th data point belonging to the $j$th class. $\Omega(f_{k})$ is an extra term specific to \texttt{xgboost} which acts as a regularisation term for the $k$th tree ($f_{k}$) used to construct the model \citep[see][for more details]{2016arXiv160302754C} which essentially penalises the size of the tree.

There are a number of hyperparameters in \texttt{xgboost} that can affect the performance of the classifier and need to be optimised, such as the size of each tree. This is performed using the \texttt{hyperopt} package \citep{pmlr-v28-bergstra13} (see Appendix~\ref{sec:hyperparams} for more details). Once the hyperparameter values are chosen, the classifier is trained according to a 10-fold stratified cross-validation scheme to choose the optimal number of trees (up to a limit of 1000). The stratification of the cross-validation folds ensures that there is the same relative number of stars in each class in the training and validation sets. At each step a new tree is added to the model and the loss function is evaluated on both a training set and validation set defined by the cross-validation fold. The training is stopped when the loss of validation set no longer decreases and starts to increase (signifying overfitting). The optimally trained model is taken as the last iteration with a decreasing validation set loss.

\section{Results}

We train the classification algorithm on the data sets of different lengths and assess the accuracy (i.e. the percentage of stars correctly classified) and reliability of each case. The classifier produces a probability that the star belongs to each class. It is common to select the final class labels as the class with the largest probability, which is the same as choosing a probability threshold of $1/C$ where $C$ is the number of classes\footnote{If more than one class exceeds the chosen threshold we assign the class with the highest probability.}. However, in this work we opt to tune the threshold to produce the largest number of true positives and lower number of false positives for the trained classifier. The details of this are explained later in this section. The classification accuracy of the classifier for each dataset length is shown in Table~\ref{tab:accuracy}. We achieve greater than 91\% accuracy across all the different lengths of time-series. The quoted accuracies take into account the three classes we want to classify the data into.

It is important not to disregard the class probabilities as they can provide useful information about the star being classified that is not apparent from the label alone. If the class probabilities are very similar then this means that the star could reasonably belong to any class, which would not be known if only the label is taken into account. In addition, the probabilities themselves can be used in subsequent probabilistic analyses as priors on the evolutionary state, for example in the classification effort being undertaken by the TESS Asteroseismic Consortium (TASC; Tkachenko et al. in prep) or in studies of the red clump \citep[e.g.][]{10.1093/mnras/stz1092,2019A&A...628A..35K}.

\subsection{Model evaluation}

\begin{figure*}
\centering
\subfloat[]{\includegraphics[width=0.45\textwidth]{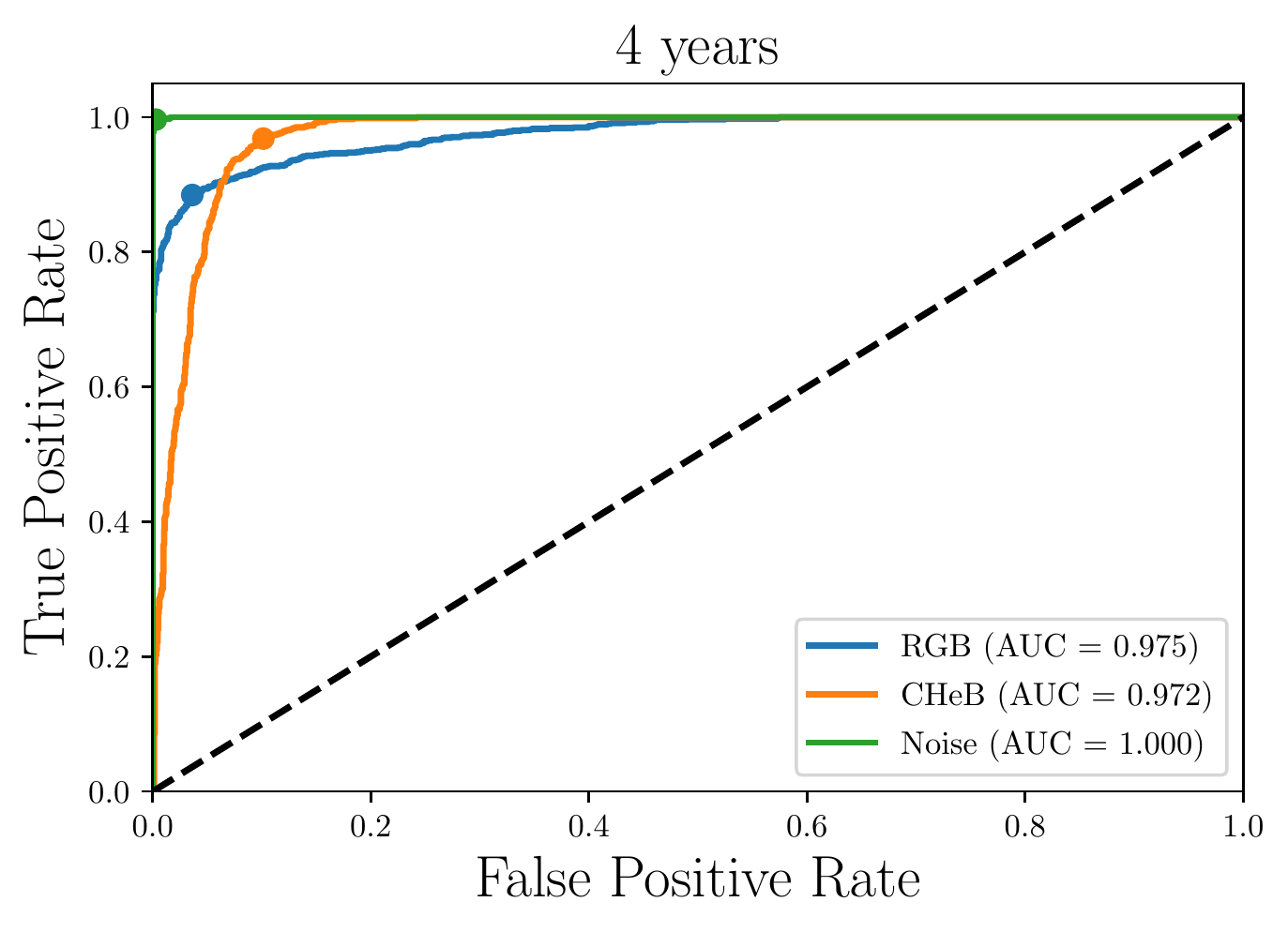}}
\subfloat[]{\includegraphics[width=0.45\textwidth]{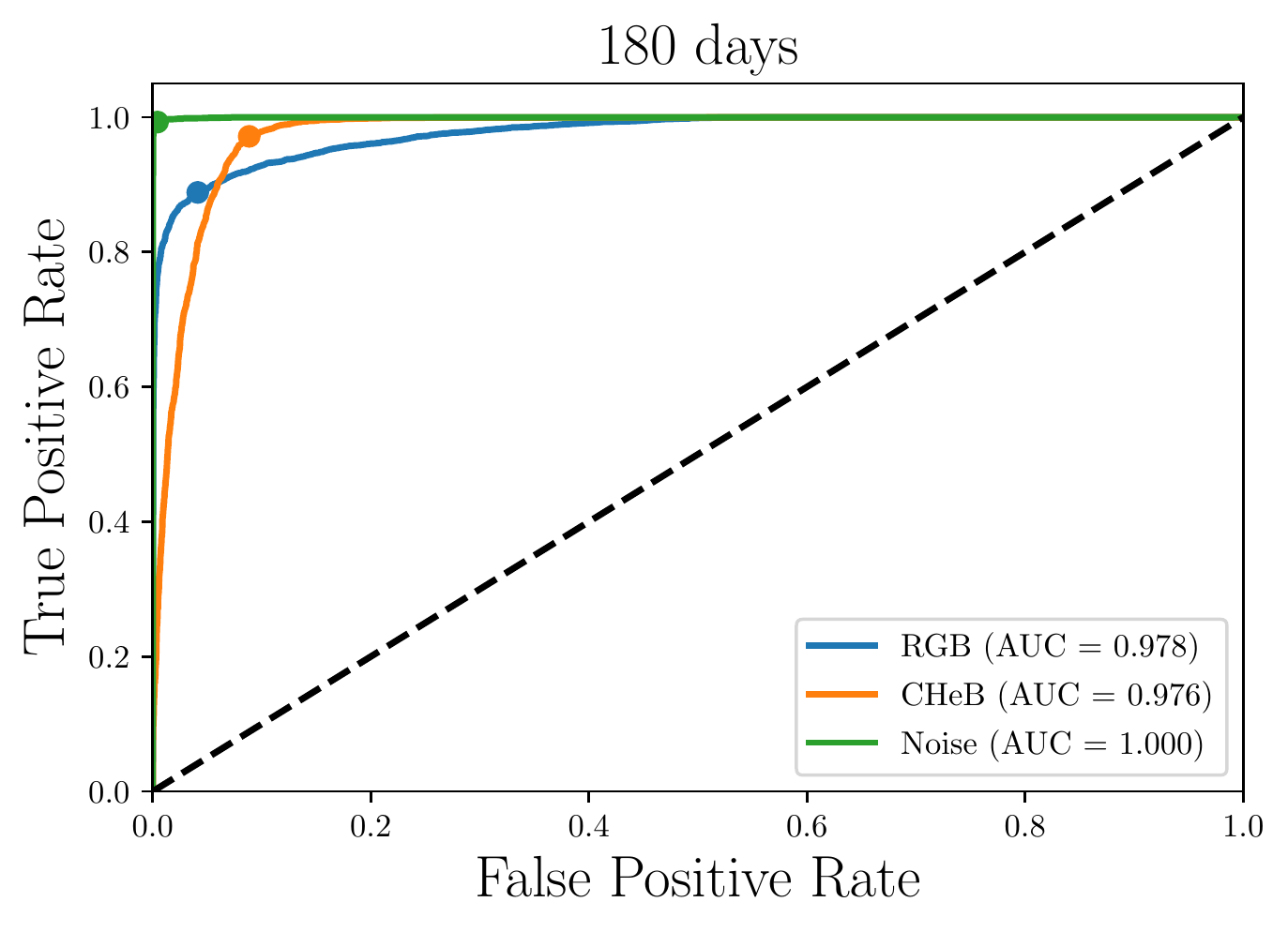}}
\\
\subfloat[]{\includegraphics[width=0.45\textwidth]{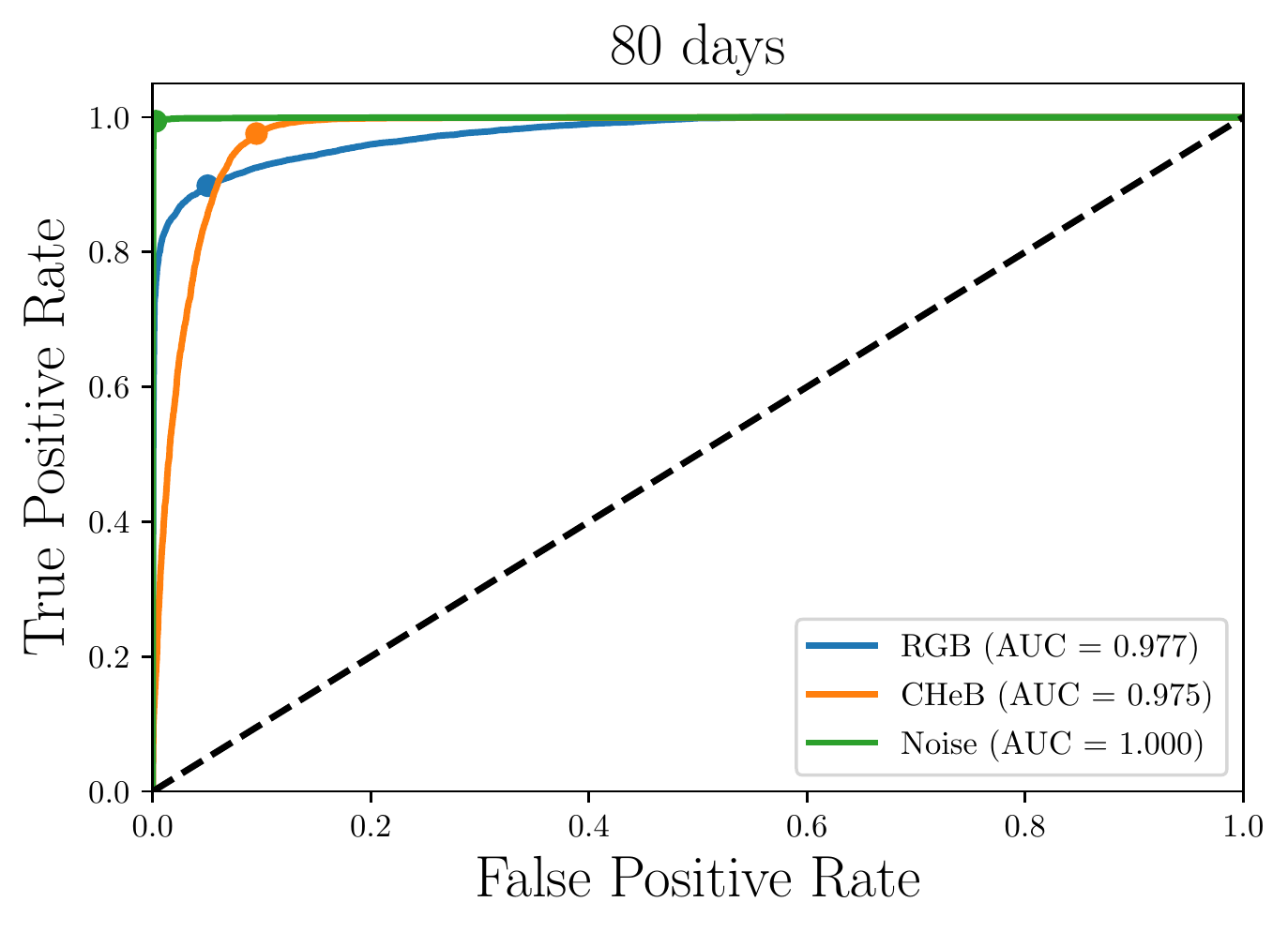}}
\subfloat[]{\includegraphics[width=0.45\textwidth]{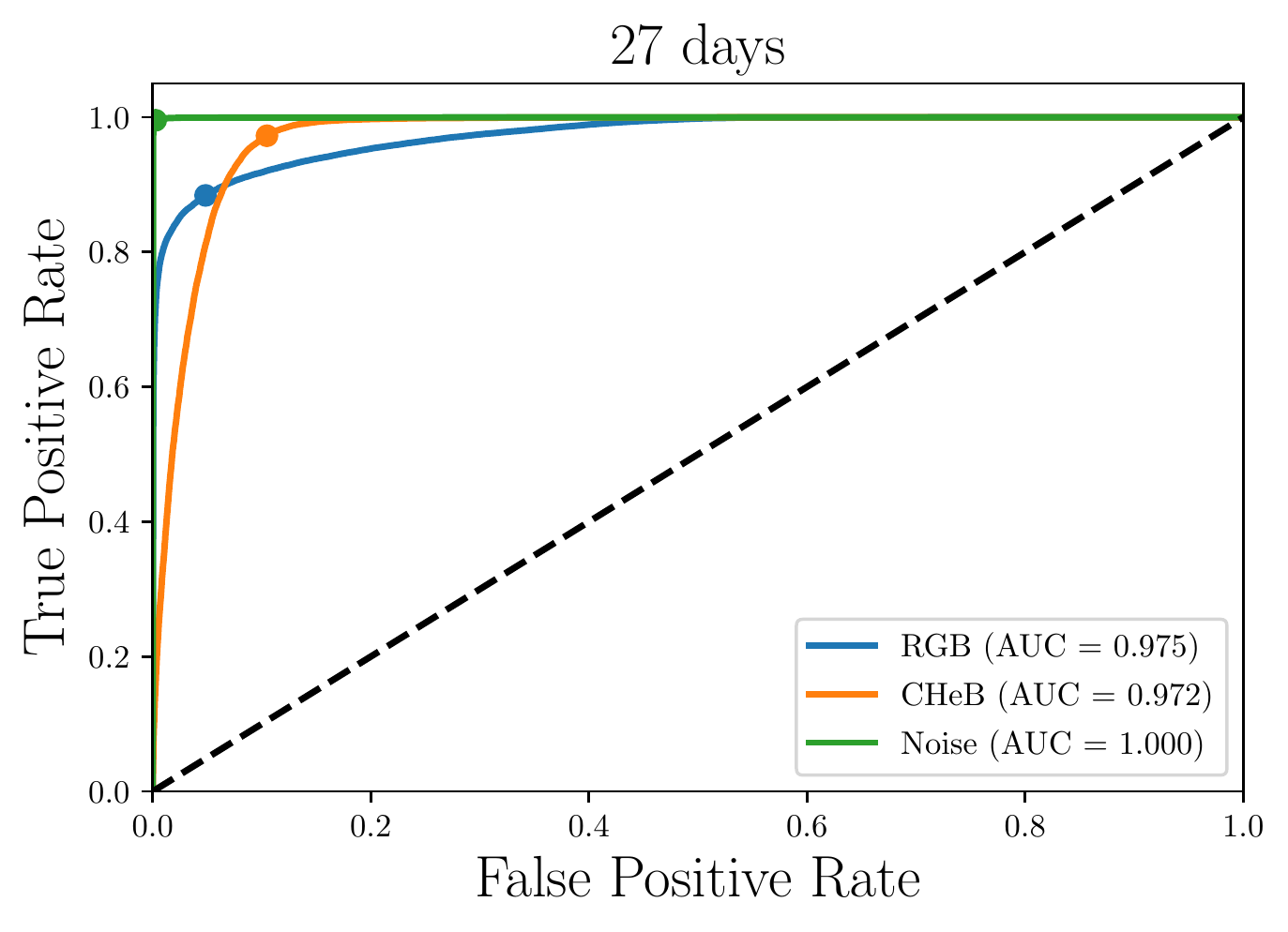}}

\caption{Receiver-operator characteristic (ROC) curves for each of the different time-series lengths. The AUC (see text) for each curve is indicated in the legend and the coloured dots show the false positive and true positive rates for the chosen probability threshold. The dashed line indicates the 1-to-1 relation.}
\label{fig:roc}
\end{figure*}

The performance of the trained models can also be assessed using the receiver-operator characteristic (ROC) curve which shows the diagnostic ability of a classifier. The curve is built up by computing the number of true and false positives in a given class for a number of different probability thresholds (decreasing from 1 to 0). The true positive rate is plotted against the false positive rate which gives us the ability to see how well the classifier correctly predicts the class for different thresholds. The ROC curves are shown for each class (using a one vs. rest methodology, e.g. \citealt[][]{bishop:2006:PRML}) and each dataset length in Fig.~\ref{fig:roc}. The probability threshold for a star to belong to a certain class is chosen using the threshold that maximises Youden's J statistic \citep[defined by the difference between the true positive rate and the false positive rate for a given threshold]{doi:10.1002/1097-0142(1950)3:1<32::AID-CNCR2820030106>3.0.CO;2-3}. The probability thresholds are therefore different for each class, reflecting the differing ability of the classifier to infer certain classes. The probability thresholds used are given in Table~\ref{tab:thresholds} and indicated by the coloured points in Fig.~\ref{fig:roc}.

\begin{table}
	\centering
	\caption{Classification accuracy of the validation set for each time-series length with and without the inclusion of $M_{\mathrm{K}_{s}}$ as a feature.}
	\label{tab:accuracy}
	\begin{tabular}{lll} 
	\hline
	Length of dataset & Accuracy (\%) & Accuracy (\%) without $M_{\mathrm{K}_{s}}$\\
	\hline
	4 years & 92.6 & 87.2\\
	180 days & 91.2 & 87.1\\
	80 days & 91.2 & 87.2\\
	27 days & 91.3 & 86.5\\
	\hline
	\end{tabular}
\end{table} 

\begin{figure*}
\centering
\subfloat[]{\includegraphics[width=0.45\textwidth]{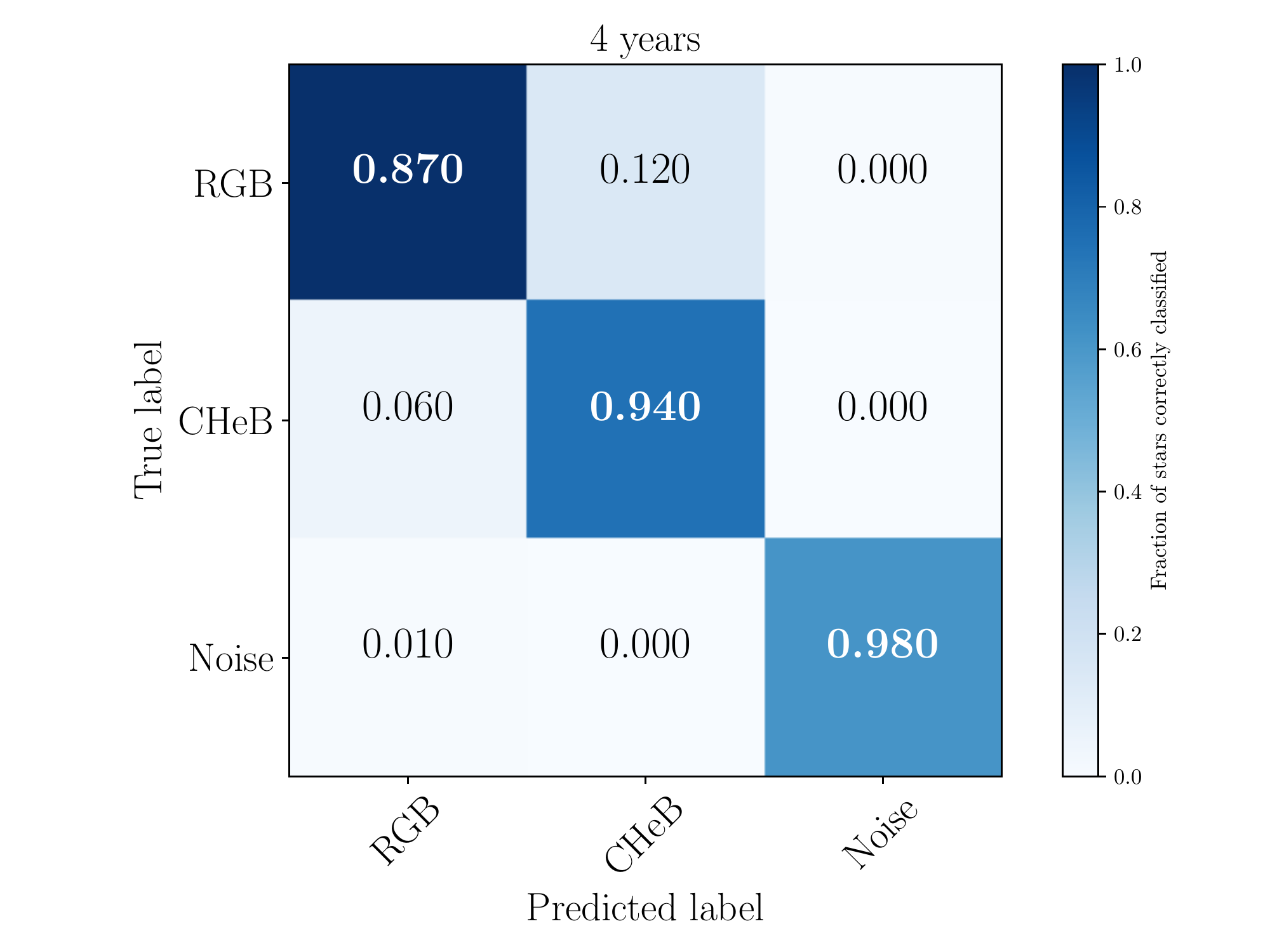}}
\subfloat[]{\includegraphics[width=0.45\textwidth]{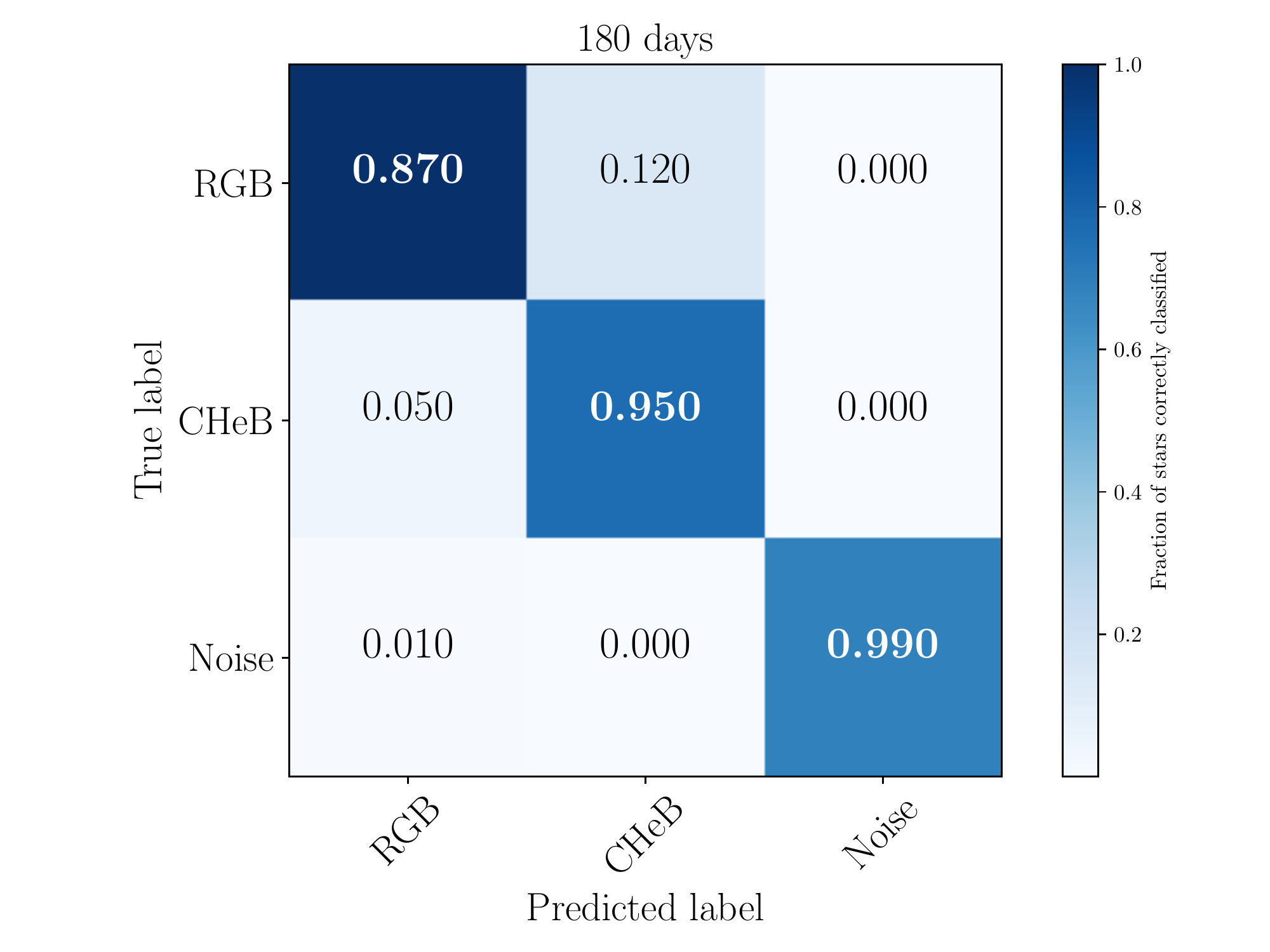}}
\\
\subfloat[]{\includegraphics[width=0.45\textwidth]{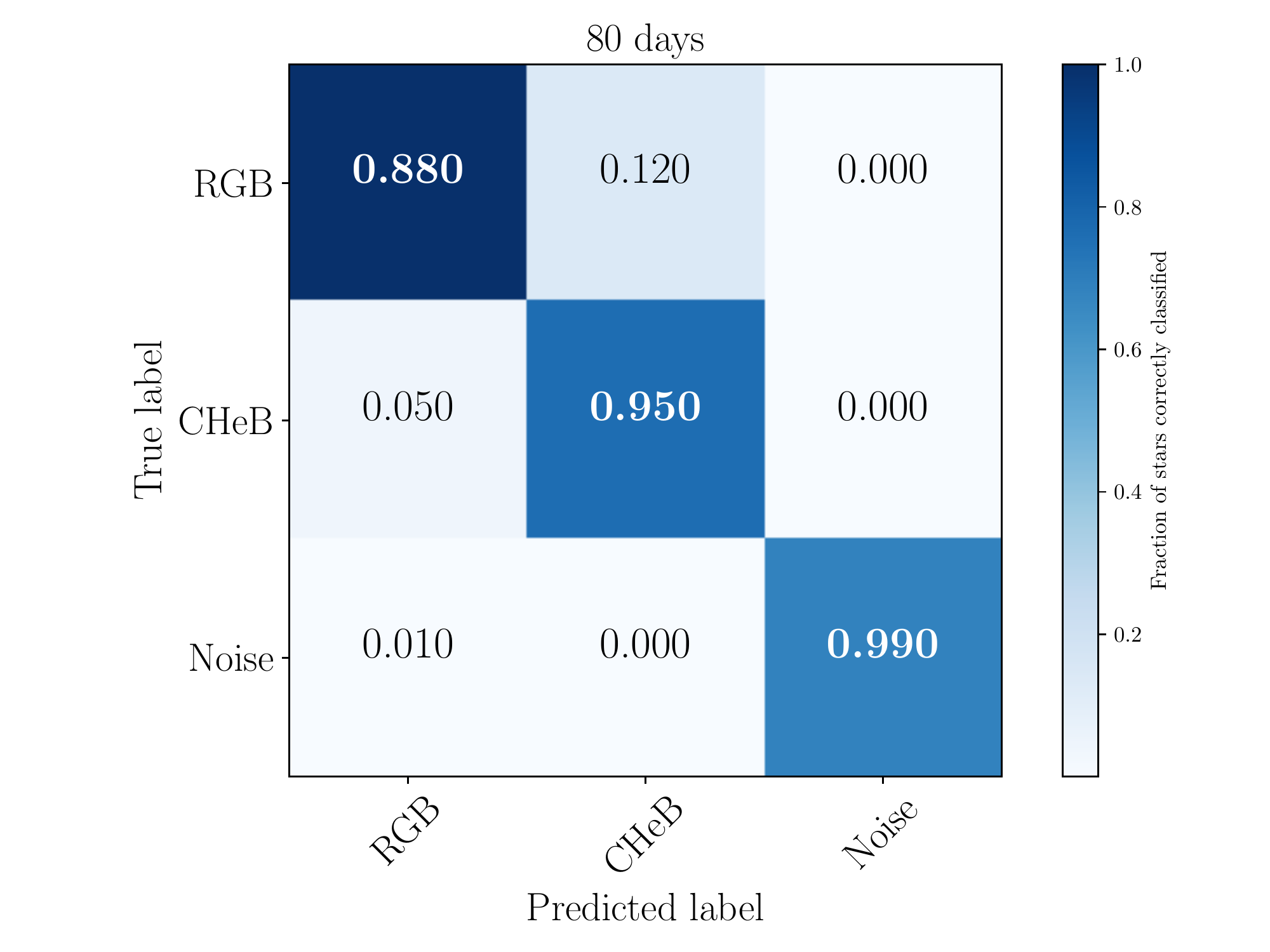}}
\subfloat[]{\includegraphics[width=0.45\textwidth]{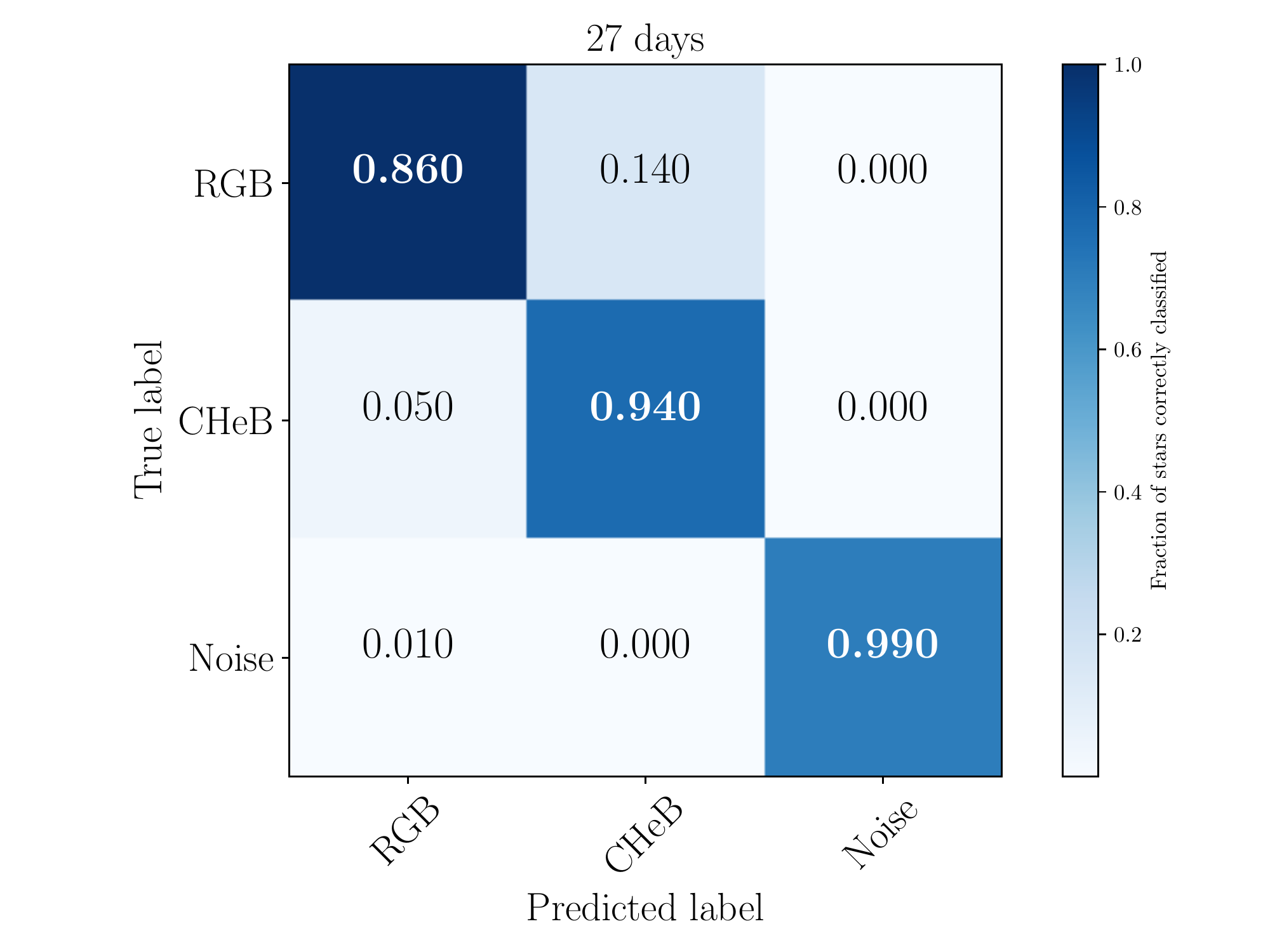}}
\caption{Confusion matrices for the classifier over each times-series length, starting with 4 years in panel (a) to 27 days in panel (d).}
\label{fig:27day_conf}
\end{figure*}

Ideally, the ROC curve should be as close to the top left corner as possible, showing that there is a threshold for which the classifier makes overwhelmingly accurate predictions with very few false positives. In our case the ROC curves show that for the Noise class the curve is very close to the ideal top left corner and so for each time-series length the classifier is able to infer this class very well, with a very low number of false positives. For the CHeB and RGB classes the performance is also very good, although we can see the interaction between the two classes manifesting itself as a reduction in the increase of the true positive rate above a value of 0.8. This provides important information about the false positive rates for the two classes. The fact that the CHeB class true positive rate increases faster than the RGB class shows that the CHeB class contains RGB false positives (as can be seen in Fig.~\ref{fig:27day_conf}), which is something we expect given the overlapping of the classes in feature space. Whilst the RGB class will also contain CHeB false positives there are far fewer of these cases.

The area-under-the-curve (AUC) of the ROC curve gives the probability that the classifier will assign a higher probability to a random sample drawn from a chosen class (i.e. the class the AUC is computed for) compared to a random sample from a different class. The higher the AUC, the better the classifier is at predicting the correct class. For all of our classifiers these values, given in the legends of Fig~\ref{fig:roc}, are close and show that the overall classification for all the classes is very good. The confusion region where the RGB stars and CHeB stars overlap is what causes the AUC values for the RGB and CHeB classes to be slightly lower than the AUC value for the Noise class. 

\begin{table}
	\centering
	\caption{Probability thresholds for a given class as a function of dataset length.}
	\label{tab:thresholds}
	\begin{tabular}{lccc} 
	\hline
	& \multicolumn{3}{c}{Class}\\
	Length of dataset & RGB & RC & ``Noise''\\
	\hline
	4 years & 0.385 & 0.493 & 0.427\\
    180 days & 0.395 & 0.283 & 0.378\\
    80 days & 0.420 & 0.307 & 0.102 \\
    27 days & 0.393 & 0.349 & 0.145 \\
	\hline
	\end{tabular}
\end{table} 

In addition to the accuracy of the trained classification models we also look at the confusion matrices to gain a sense of which classes the classifier predicts correctly and which provide the most difficulty, as shown in Fig.~\ref{fig:27day_conf}. The diagonal of the confusion matrix shows the proportion of stars in a given class that have been correctly classified and the off-diagonal elements show how a star of a given class has been misclassified. Looking at Fig.~\ref{fig:27day_conf} it is apparent that the main source of confusion is between the CHeB class and the RGB class. This is expected as in the confusion region the two classes overlap which will cause confusion and results in a drop in accuracy.

\subsection{Model intepretation}

\begin{figure*}
\centering
\subfloat[]{\includegraphics[width=0.45\textwidth]{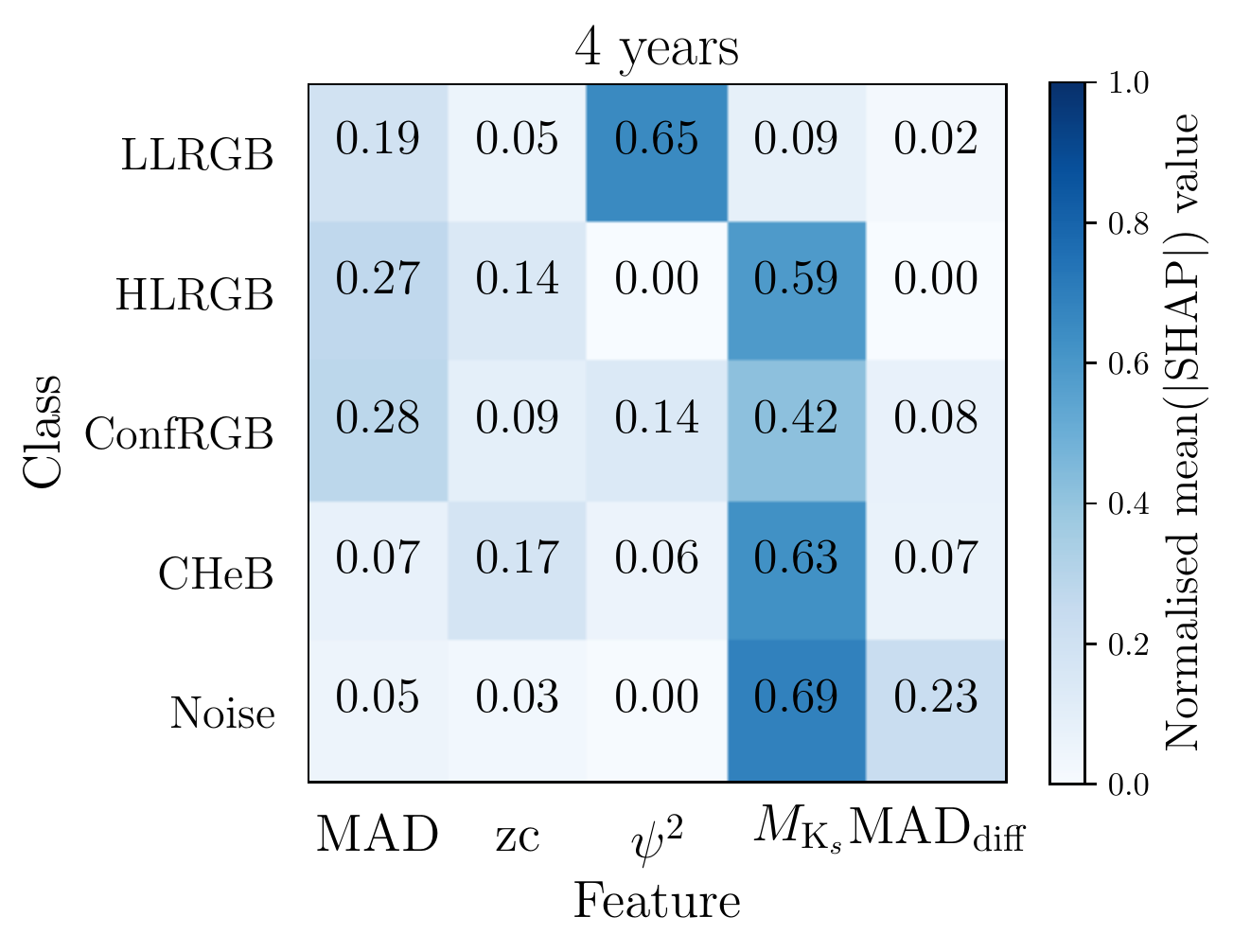}}       
\subfloat[]{\includegraphics[width=0.45\textwidth]{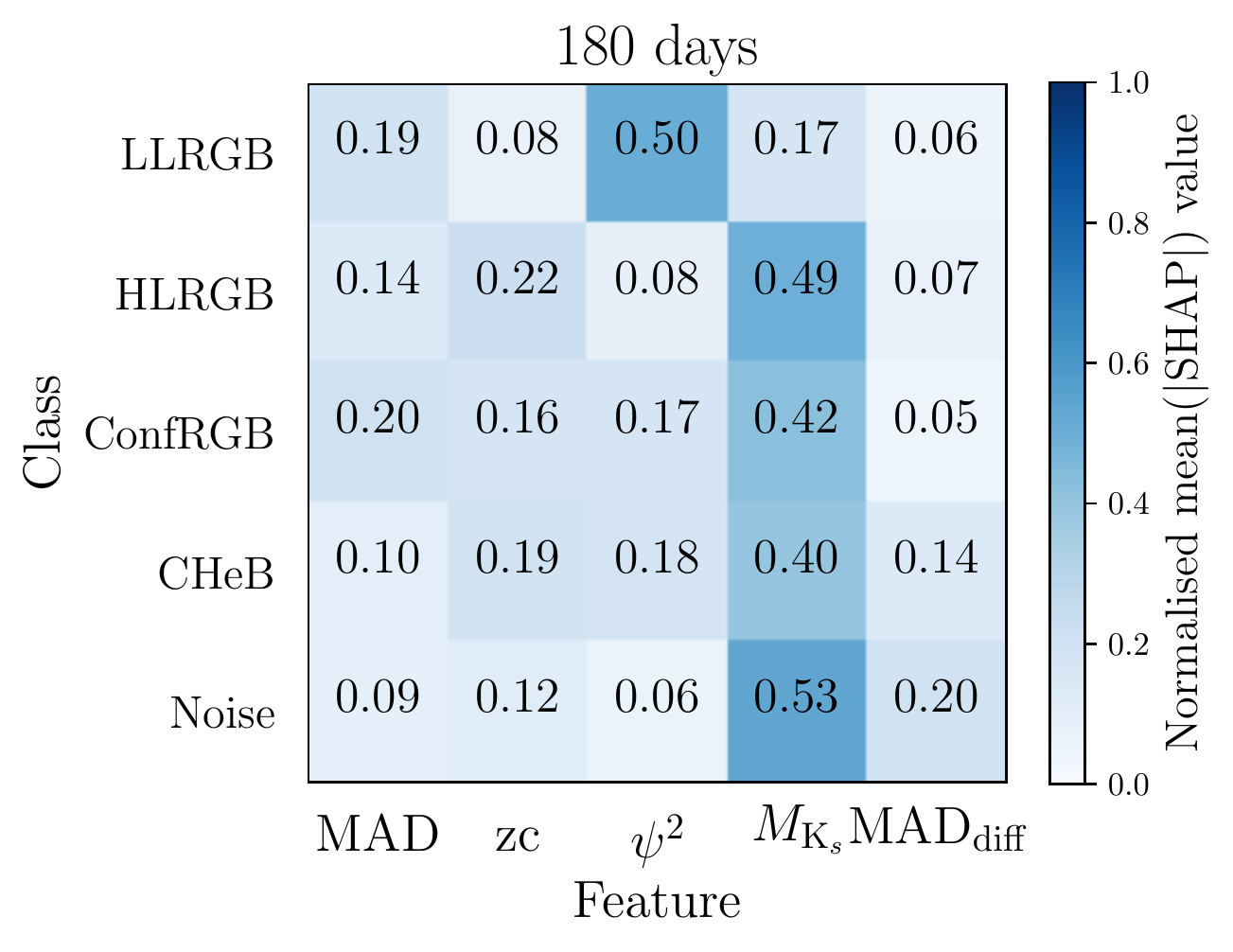}}
\\
\subfloat[]{\includegraphics[width=0.45\textwidth]{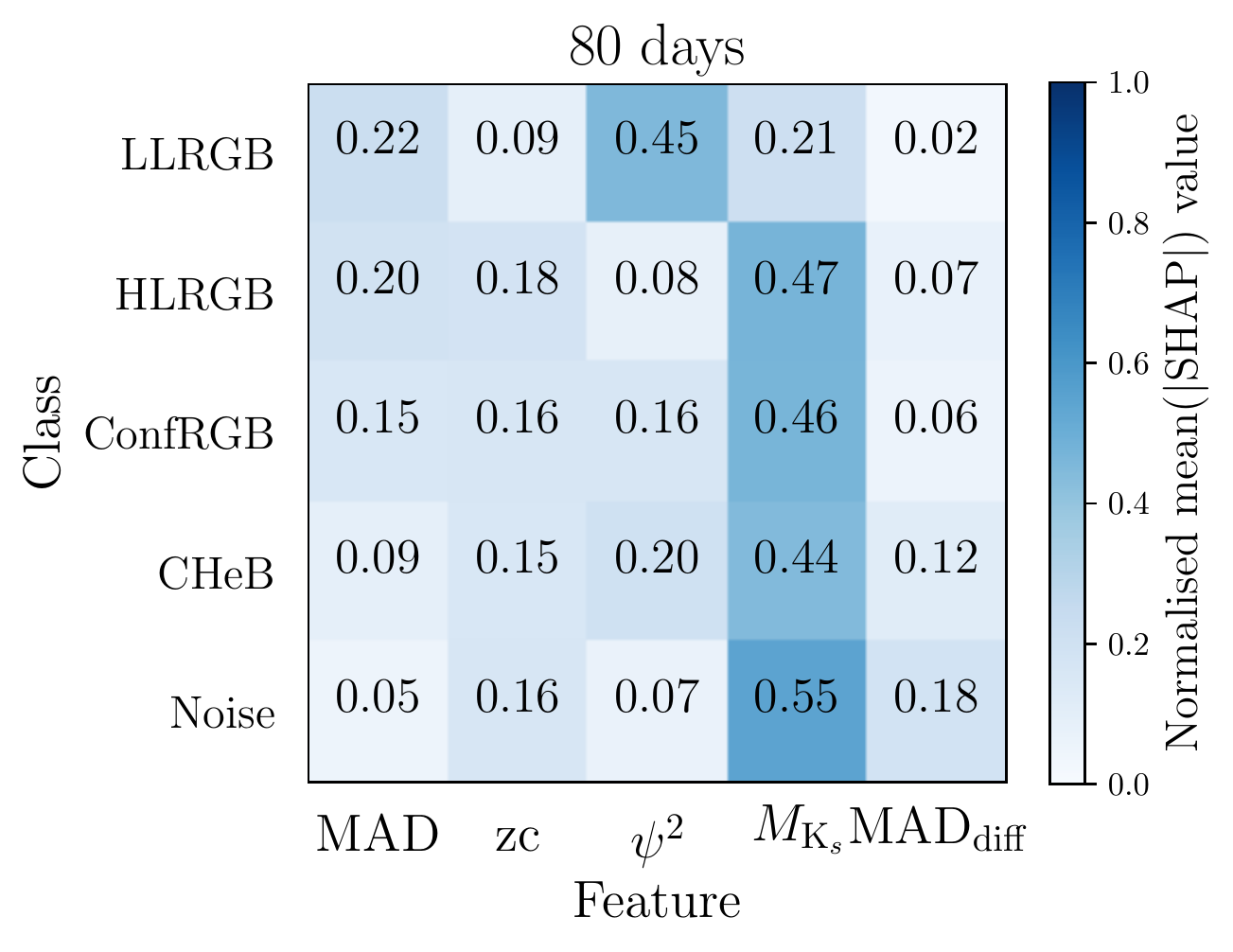}} 
\subfloat[]{\includegraphics[width=0.45\textwidth]{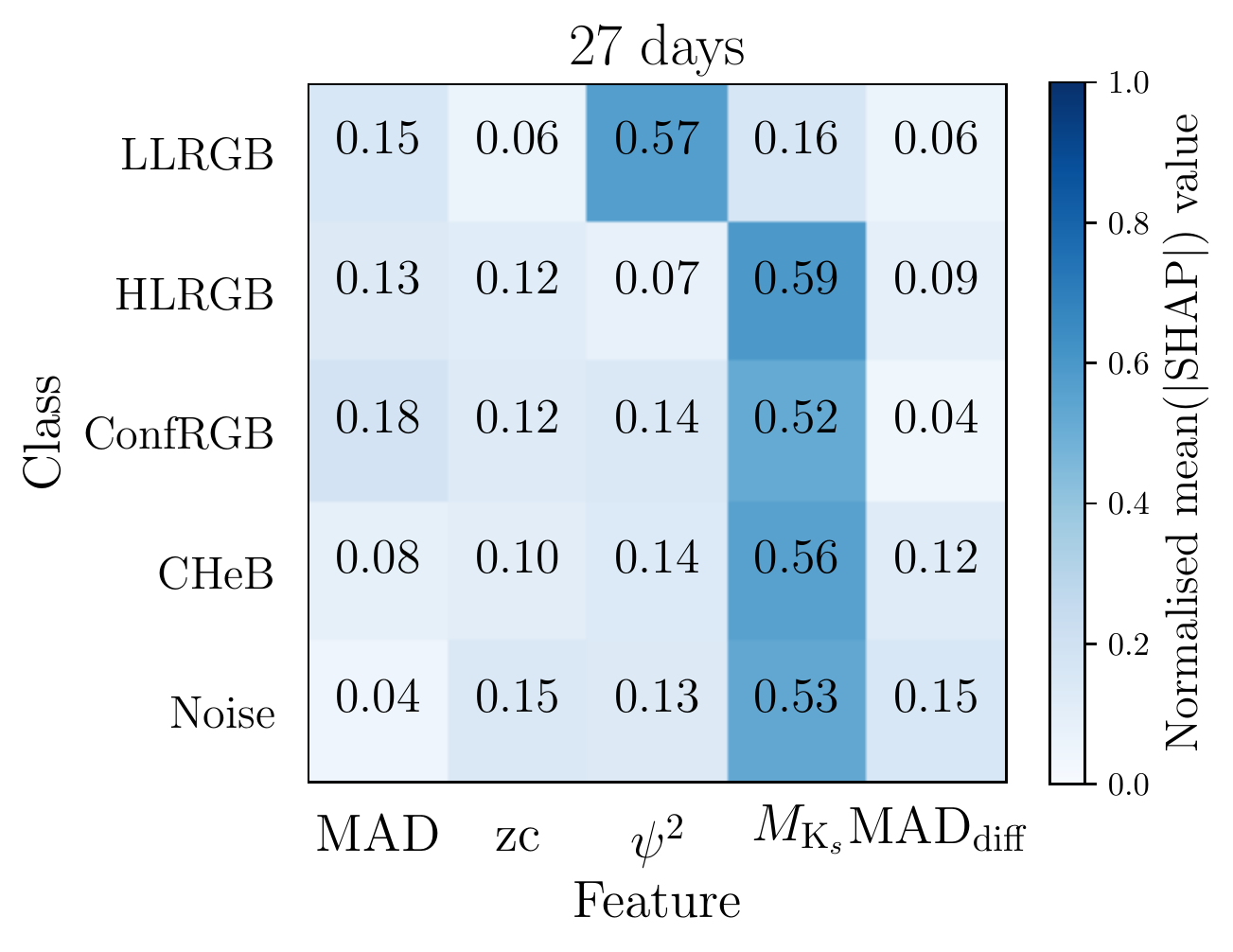}} 
\caption{Feature matrices for each trained model, arranged according to descending timeseries length. The normalised number of zero-crossings are labelled as ``zc''.}
\label{fig:feature_matrix}
\end{figure*}

The interpretability of the trained model is key to understanding how it performed and, more importantly, why it is performing in such a way. We use the feature importance of the classifier to interpret the model and which features it uses to classify stars belonging to different classes. We choose to use SHapely Additive exPlanation (SHAP) values \citep[see][for a more in-depth overview]{lundberg2018consistent, NIPS2017_7062} to provide feature importances. The feature importance information is presented in Fig.~\ref{fig:feature_matrix} as a ``feature matrix'', which contains the SHAP value averaged over each star in a given class normalised such that the rows sum to 1. Each row of the matrix then tells us the feature importance for the class represented by that row, and the column shows us the relative feature importance between each class.

 We do not necessarily expect the same ranking of features for each length of dataset since differences in the dataset length may have an effect on the ability to derive given features. For example the shorter the dataset the more likely the variance is to differ from realisation to realisation due to the stochastic nature of the data. Across the different lengths of data $M_{\mathrm{K}_{s}}$ is the most important feature for the majority of classes, which is to be expected. Whereas for the LLRGB stars the most important feature is $\psi^{2}$ because this provides the best way to distinguish between the nearby Noise class. Therefore, it is important to consider a local treatment of feature importance, otherwise a feature may be disregarded when looking globally when in fact it is vital to the accurate classification of a specific class.

\section{Application to the Full \emph{Kepler} long-cadence dataset}\label{sec:fullkep}

The sub-sample of data that we are looking at in this work, red giants showing solar-like oscillations, may not give us a full insight into feature space. Since all of our stars are stochastic oscillators, the underlying process generating the oscillations is the same. In order to interpret the features fully it is helpful to expand the region of parameter space with more stars with different types of intrinsic variability (e.g. activity and stellar oscillations) and extrinsic variability (e.g. eclipsing binaries). This can help identify contaminants in future samples. For instance, a star that has more than one type of variability, e.g. solar-like oscillations and eclipses, or this can be used to classify a wider range of stellar variability types. To help with our interpretation we opt to extract features from every object observed in long cadence with \emph{Kepler}. Since we will now have many different types of objects, e.g., eclipsing binaries, classical pulsators etc., this will not only enable us to improve the interpretation of our features, but also show the viability of using the derived features to discern other variable classes of star.

\begin{figure*}
\centering
\subfloat[\label{fig:fullkep}]{\includegraphics[width=\columnwidth]{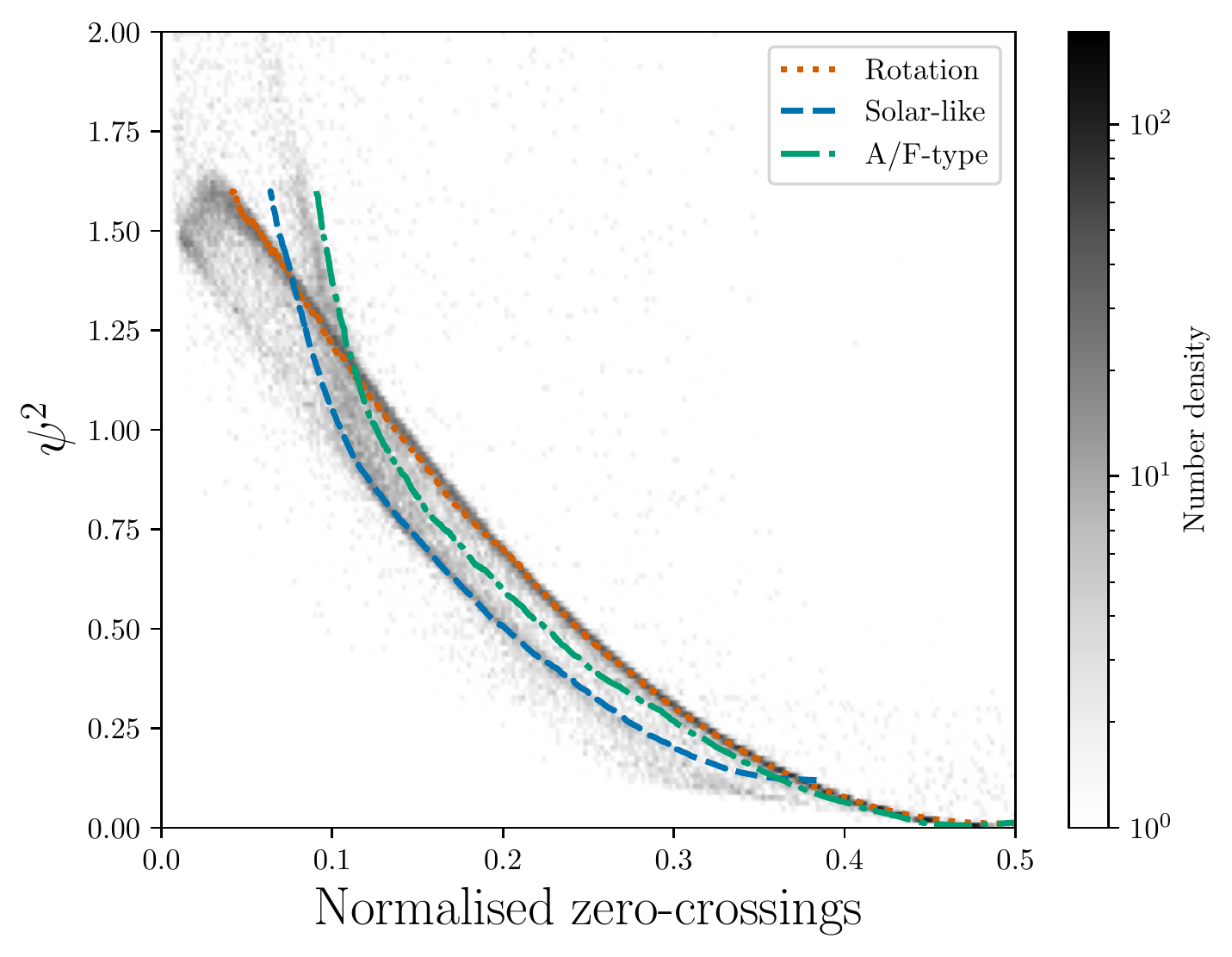}}
\subfloat[\label{fig:fullkep_full}]{\includegraphics[width=\columnwidth]{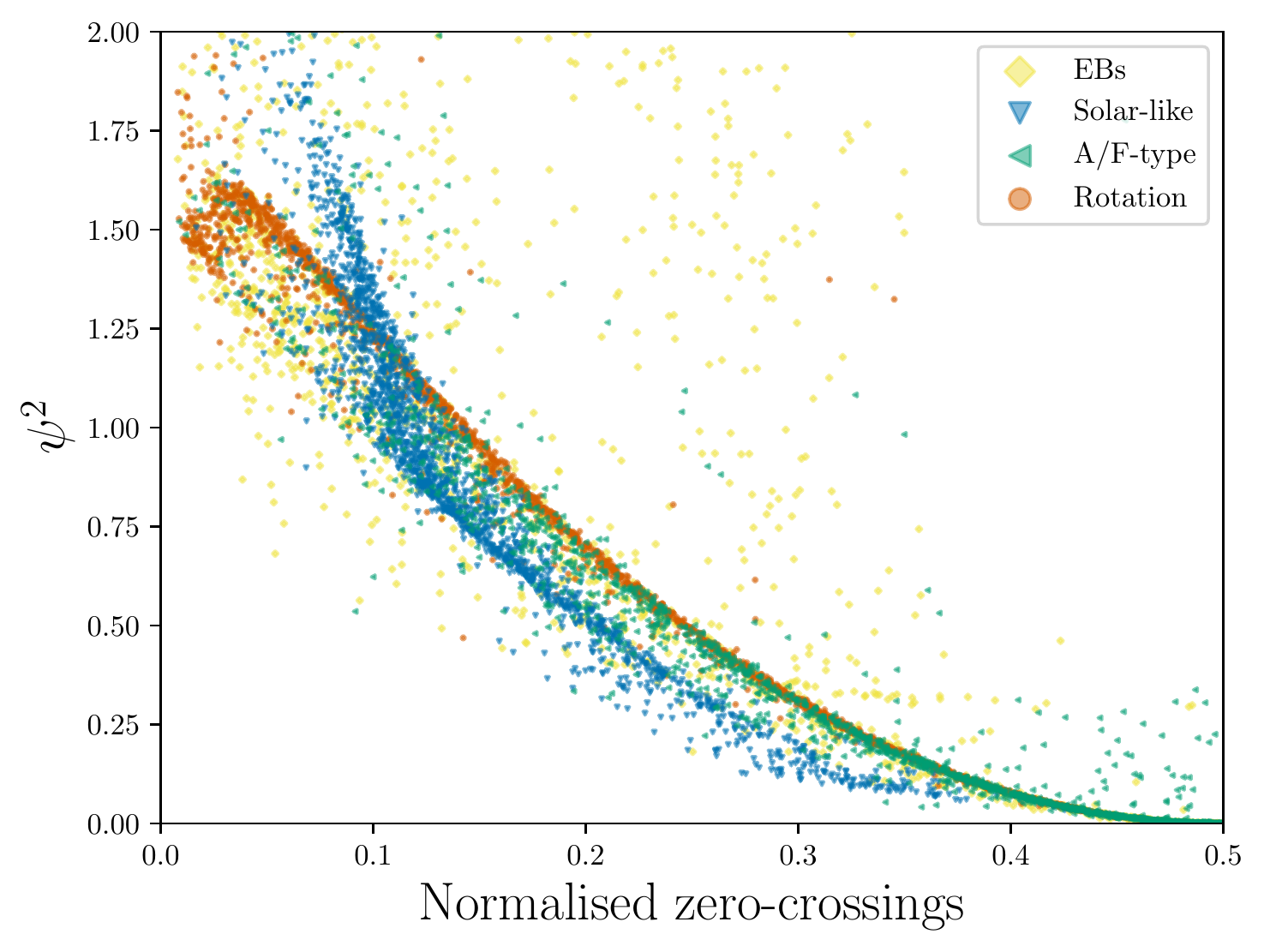}}
\caption{Panel (a) shows a hexbin plot showing the $\psi^{2}$ feature (coherency) as a function of the normalised zero-crossings for the full \emph{Kepler} field, coloured by number density. Overplotted are the approximate positions of stars showing rotational-modulation, solar-like oscillations and $\delta$-Scuti/$\gamma$-Doradus stars (labelled as A/F-type main-sequence stars) computed using a rolling mean of width 0.05 in normalised zero-crossings. Panel (b) shows the distribution of a random sample of 2000 stars in each category over the same region of parameter space as panel (a) with the addition of eclipsing binaries (EBs).}
\label{fig:fullkep}
\end{figure*}

Let us focus on two parameters in particular, the number of normalised zero-crossings and the coherency parameter $\psi^{2}$. These two parameters provide information regarding the timescale of the dominant contribution to the lightcurve and the degree of coherency (or stochasticity). Fig.~\ref{fig:fullkep} shows these features for the full \emph{Kepler} sample in a reduced region of parameter space, which we shall explain before moving to the larger view. There are two visible strands that extend with decreasing zero-crossings and increasing coherency, which are indicative of distinct populations. The strands cross at very long timescales (i.e a low number of normalised zero crossings), with the lower strand crossing over the more coherent, higher strand at $\sim0.1$ in the normalised number of zero-crossings. 

The visible separation leaves the impression that stars with different properties can lie on each of these strands and so we overplot the populations of known types to investigate where they fall in parameter space. This is shown in Fig~\ref{fig:fullkep_full}. We include stars showing rotation from \cite{2014ApJS..211...24M}, solar-like oscillators from \cite{2019arXiv190300115H} which have a detection probability of 1, eclipsing binaries from the Kepler Eclipsing Binary Catalogue \citep{2016AJ....151..101A,2016AJ....151...68K,2011AJ....141...83P} and a sample of A and F-type stars that lie in the $\delta$-Scuti instability strip that include $\delta$-Scuti and $\gamma$-Doradus variables from \cite{2019MNRAS.485.2380M}.

The higher, more coherent strand is populated by stars exhibiting some degree of rotational modulation and the lower, less coherent strand is in fact populated by solar-like oscillators. This conforms with expectations, since solar-like oscillators are inherently stochastic and their oscillations are mostly much less coherent than signal from rotational modulation, and so for any given timescale (inferred by proxy from the number of zero-crossings) a solar-like oscillator and rotational variable can be distinguished through the coherency of the signal. However, the strands do cross at a value of ~0.1 in normalised zero-crossings. This is due to the fact that the granulation time-scale is closely linked to the radius of the star, the larger the radius of the star, the larger the convective cells which leads to a larger granulation timescale, i.e. a decreasing number of zero-crossings. The rate of ascent of a star up the red-giant branch is not linearly proportional to the number of zero-crossings. Therefore, we interpret this sudden rapid increase in the coherency of the signal as due to the increasingly fast evolution of the star towards the tip of the red-giant-branch.

Eclipsing binaries appear in Fig~\ref{fig:fullkep_full} in two different configurations depending on the dominant contribution to the signal in the data. If the dominant contribution is stellar in origin, i.e. pulsations, then the star will appear on the strands following the stellar signal. However, if the dominant contribution is instead from the eclipses then they will lie away from the strands with a high coherency corresponding to the regularity of the eclipse signal and a normalised zero-crossing value proportional to the binary period. 

Finally, we have the classical pulsators which appear at two different extremes of Fig~\ref{fig:fullkep} due to the different pulsations observed. $\gamma$ Dor pulsators are gravity-mode pulsators (where buoyancy is the restoring force of the pulsations) and so oscillate at long periods, this is evident by the low normalised number of zero-crossing values and high coherency of the signal. Whereas $\delta$ Sct pulsators are pressure-mode pulsators (where pressure is the restoring force of the pulsations) and oscillate at much higher frequencies placing them at the high extreme in normalised zero-crossings. Their signals in long-cadence are seen to be less coherent than the $\gamma$ Dor pulsators, most likely due to the fact that they oscillate above the long-cadence Nyquist frequency (of $\sim 24$day$^{-1}$) and so it is likely that the lack of coherency comes from the under-sampling of the data rather than from the oscillation modes themselves.

The investigation of the timeseries features chosen in section \ref{sec:features} applied to the full \emph{Kepler} dataset shows that not only can these features distinguish RGB from CHeB stars, but they could also potentially be used to classify a wider range of variable stars.

\section{Discussion}

The trained classifiers perform very well over each dataset length, however there are a few assumptions that we have made that should be addressed. Like any machine-learning task, the ability to generalise to unseen data can be affected by the quality of the training data. There could be an issue if the underlying populations between the new data and the training data differ significantly. We are therefore making the assumption that the underlying population observed by \emph{Kepler} would be close to other observed underlying populations, for example, observed by K2 or TESS. Given that \emph{Kepler} stared at a single patch of sky for 4 years and other missions, such as K2 and TESS, look over more and larger patches of sky this assumption may not be strictly valid. This also applies to the properties of the particular mission. Time-series that this classifier can be applied to must have a cadence that is similar to the training data. Otherwise this will introduce biases into the computed features and lead to incorrect classification. The feature that the differences between the training data and new unseen data have the largest effect on is the K-band absolute magnitude, whereby population level effects are ignored in its calculation (e.g. \citealt{2001MNRAS.323..109G,2016ARA&A..54...95G}), using Eq.~\ref{eqn:absmag}. Therefore slight changes in the underlying $M_{\mathrm{K}_{s}}$ distribution could cause the predictions from the classifier to be affected. However, this is the reason why we use an intrinsic property rather than parallax and apparent magnitude, since absolute magnitude should be independent of distance. The other features could also be affected if the distribution of stellar parameters (e.g. mass, radius, metallicity and effective temperature) is significantly different from those of the \Kepler training set. This effect is easily quantifiable by comparing the training set feature distributions to those from a new data set. Secondly, we have not taken into account the difference in bandpasses between \emph{Kepler} and TESS in which granulation and oscillation signals are expected to have lower amplitudes in TESS data due to the redder bandpass \citep{2016ApJ...830..138C}. Provided that the time-series are dominated by granulation and physical signal then this can be approximately accounted for with a multiplicative factor in variance \citep{2016ApJ...830..138C}.

\begin{figure*}
\centering
\includegraphics[width=0.45\textwidth]{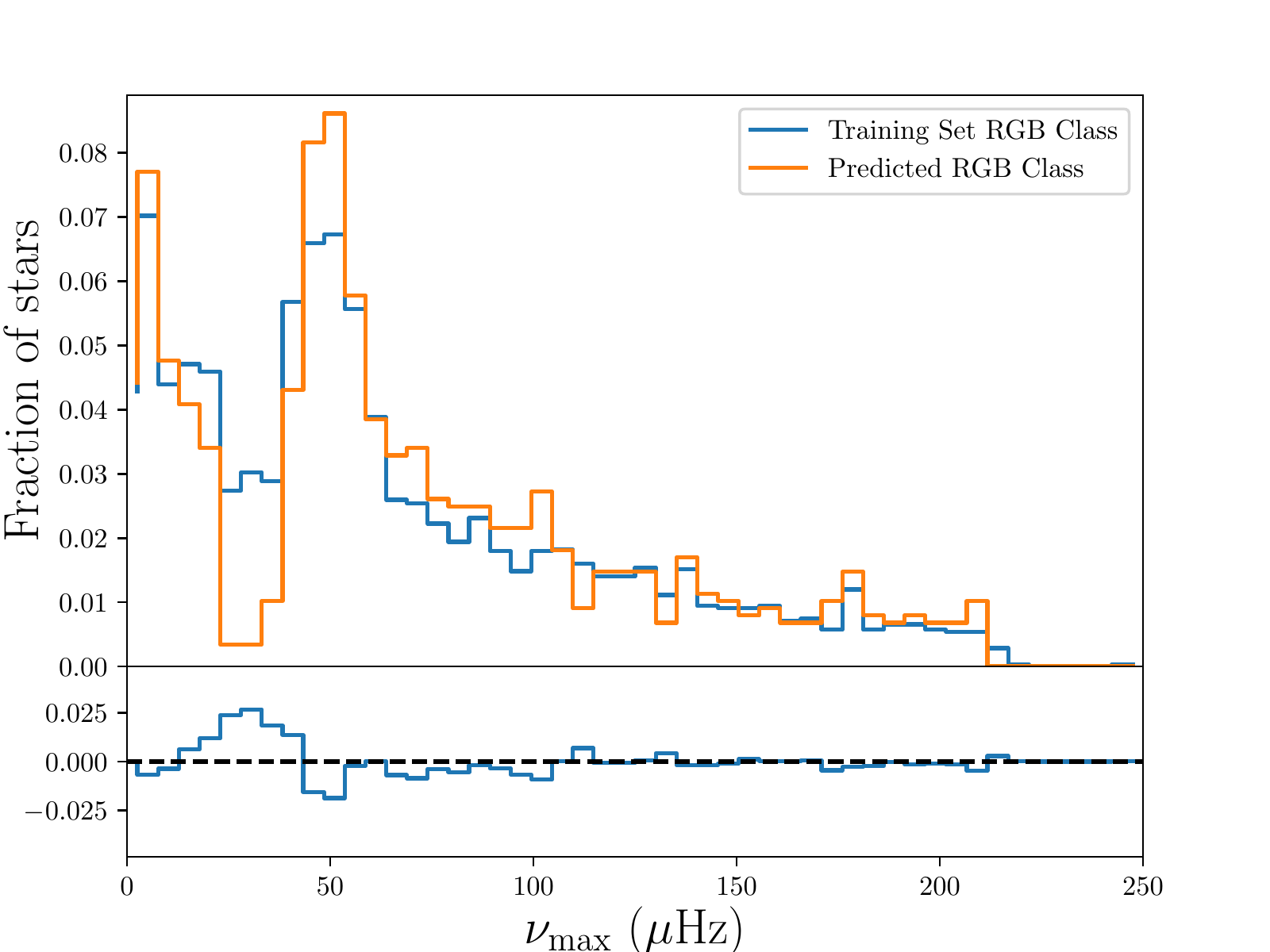}
\includegraphics[width=0.45\textwidth]{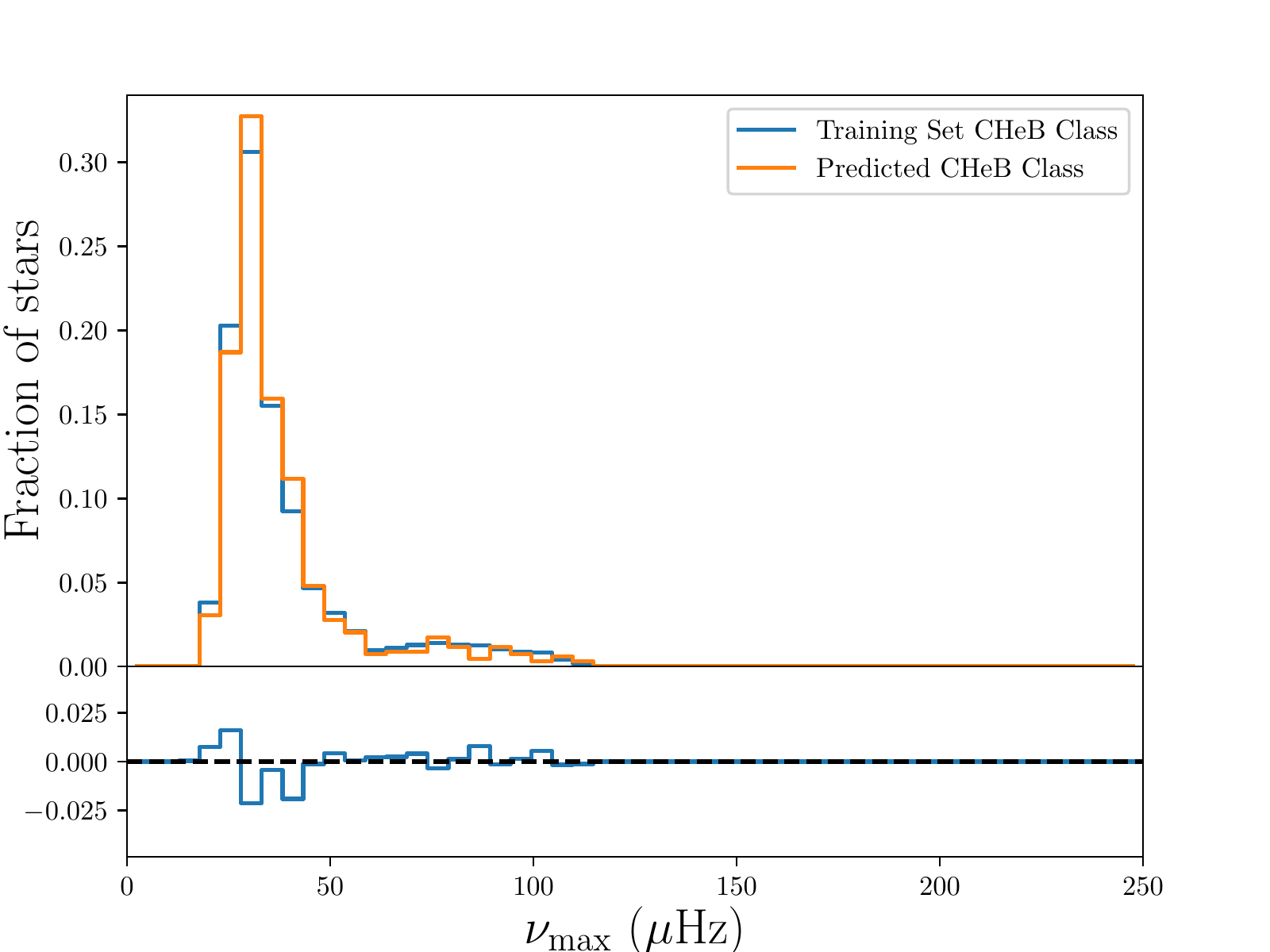}
\caption{Comparison of the distribution of RGB stars (left) and CHeB stars (right) in the training set (blue) and the predictions (orange). The difference between the training set and prediction distributions are shown below each histogram.}
\label{fig:rgb_rc_preds}
\end{figure*}

The fraction of RGB and CHeB stars in the training set compared to fraction of RGB and CHeB stars predicted are given in the two panels of Fig~\ref{fig:rgb_rc_preds}. For the CHeB stars we perform well and the distributions of stars in our training set and our predictions are very close. It can also be seen for the case of the secondary clump stars, the small hump around $\nu_{\mathrm{max}}\sim 80\mu$Hz, due to the agreement between the two distributions. Whereas for the RGB stars there is good agreement for the majority of $\nu_{\mathrm{max}}$ except for the region around the red clump. There is a clear paucity of predicted RGB stars in the region $20<\nu_{\mathrm{max}} (\mu\mathrm{Hz})<50$ which means that these RGB stars have been classified as CHeB stars. In addition, there is also a slight overabundance of RGB stars at $\nu_{\mathrm{max}} \sim 50-60\;\mu\mathrm{Hz}$ which means that CHeB stars have been misclassified as RGB stars. This is due to the class imbalance in the region around the red clump, which can be as high as 7:1 in favour of CHeB stars at $\nu_{\mathrm{max}} \sim 30\;\mu\mathrm{Hz}$.

There are two other cases that can result in a slightly poorer performance of the classifier which are not due to the classifier itself. The first is if the computed absolute magnitude is not correct, this could be due to source confusion during the cross-match to the Gaia data resulting in choosing the wrong distance estimate. In the event that the computed absolute magnitude is not correct, this will naturally cause the classifier to misclassify the star in question. This situation could arise with any of the other features, however it is most likely to occur for the absolute magnitude calculation due to source confusion. The second case is when the white noise level is very high, either due to a large amount of shot noise or because the star is very faint. As such the granulation signal can be very hard or impossible to detect which will result in the calculated features being in the wrong part of feature space. This will result in the star being classified as noise. It can to some degree be accounted for in the shorter length time-series, whereby the signal-to-noise level is lower. However, more extreme situations cannot be accounted for in our training set. 

Despite training our models on real data from \emph{Kepler}, the described accuracy is likely an upper limit when applied to other datasets, e.g. K2, TESS or, in the future, PLATO. This is due to the different way in which pipelines detrend the data, which can affect the extraction of the zero-crossings (and higher-order crossings).

In order for the probabilities returned by the classifier to be used in subsequent analyses it is important for the probabilities to be calibrated. By calibrated we mean that if the probability returned for a star is 0.7, then we would expect this to be correct 70\% of the time. This means that the probability produced by the model is representative. It is well known that tree-based methods produced calibrated probabilities and so we do not include any post-processing to do so (see Appendix~\ref{sec:prob_calib}).

\section{Conclusion}

In this work we have trained a series of classifiers on differing length datasets that can robustly and accurately classify the evolutionary state of red-giant stars. The classifiers make use of time domain features; the MAD of the time-series and its first differences, the normalised number of zero-crossings and a coherency measure $\psi^{2}$. In addition to the time-domain features the absolute K-band magnitude is also used. The trained models for each time-series length and the code used to extract the features are available at \url{https://github.com/jsk389/Clumpiness}. All classifiers, including that applied to the shortest dataset length simulating the shortest datasets available from TESS achieve above 91\% accuracy when inferring red giant evolutionary states. As a result, our classifier will be highly applicable to classifying the large number of stars observed with TESS and, in the future, PLATO as well.

We have also shown that our probabilities are well calibrated and so can be readily applied to probabilistic analyses that require specific populations of stars, e.g. the red-clump, or that can be used as prior probabilities, e.g. in future peak-bagging codes \citep[e.g.][]{2019FrASS...6...21C}.

The time-series features that we propose are not only useful for determining evolutionary states, they are also of great use for identifying different stellar populations. This has been demonstrated briefly for a random subset of all the stars observed with \emph{Kepler} and these can be of great use in subsequent classification tasks applied to multiple stellar populations. For example, both the normalised number of zero-crossings and the coherency measure $\psi^{2}$ are being used as features in the TESS Data for Asteroseismology (T'DA) classification work (Tkachenko et al. in prep).

\section*{Acknowledgements}

We would like to thank Marc Hon, Yvonne Elsworth and Nathalie Themeßl for
their very useful comments and discussions. The research leading to the presented results has received funding from the European Research Council under the European Community's Seventh Framework Programme (FP7/2007-2013) / ERC grant agreement no. 338251 (StellarAges). KJB is supported by the National Science Foundation under Award No.\ AST-1903828. This work has made use of data from the European Space Agency (ESA) mission {\it Gaia} (\url{https://www.cosmos.esa.int/gaia}), processed by the {\it Gaia} Data Processing and Analysis Consortium (DPAC, \url{https://www.cosmos.esa.int/web/gaia/dpac/consortium}). Funding for the DPAC has been provided by national institutions, in particular the institutions participating in the {\it Gaia} Multilateral Agreement. This research has made use of the NASA Exoplanet Archive, which is operated by the California Institute of Technology, under contract with the National Aeronautics and Space Administration under the Exoplanet Exploration Program.

This work made use of the \url{gaia-kepler.fun} crossmatch database created by Megan Bedell.


\section*{Data Availability}
 
The data underlying this article will be shared on reasonable request to the corresponding author.



\bibliographystyle{mnras}
\bibliography{manuscript} 



\appendix

\section{Accounting for fill in the computation of higher-order crossings}\label{sec:fill}

The presence of gaps in the time-series will affect the estimation of the number of zero crossings, and subsequent higher order differences, causing an underestimation of the underlying value of the number of zero-crossings. A correction therefore needs to be applied to account for gaps in the timeseries data so that we can extract reliable zero-crossing estimates from datasets with missing data. The purpose of such a correction should be that when applied, the number of zero-crossings estimated accounting for the fill accurately reflects the underlying number of zero-crossings if the data had no gaps.

\begin{figure}
\includegraphics[width=\columnwidth]{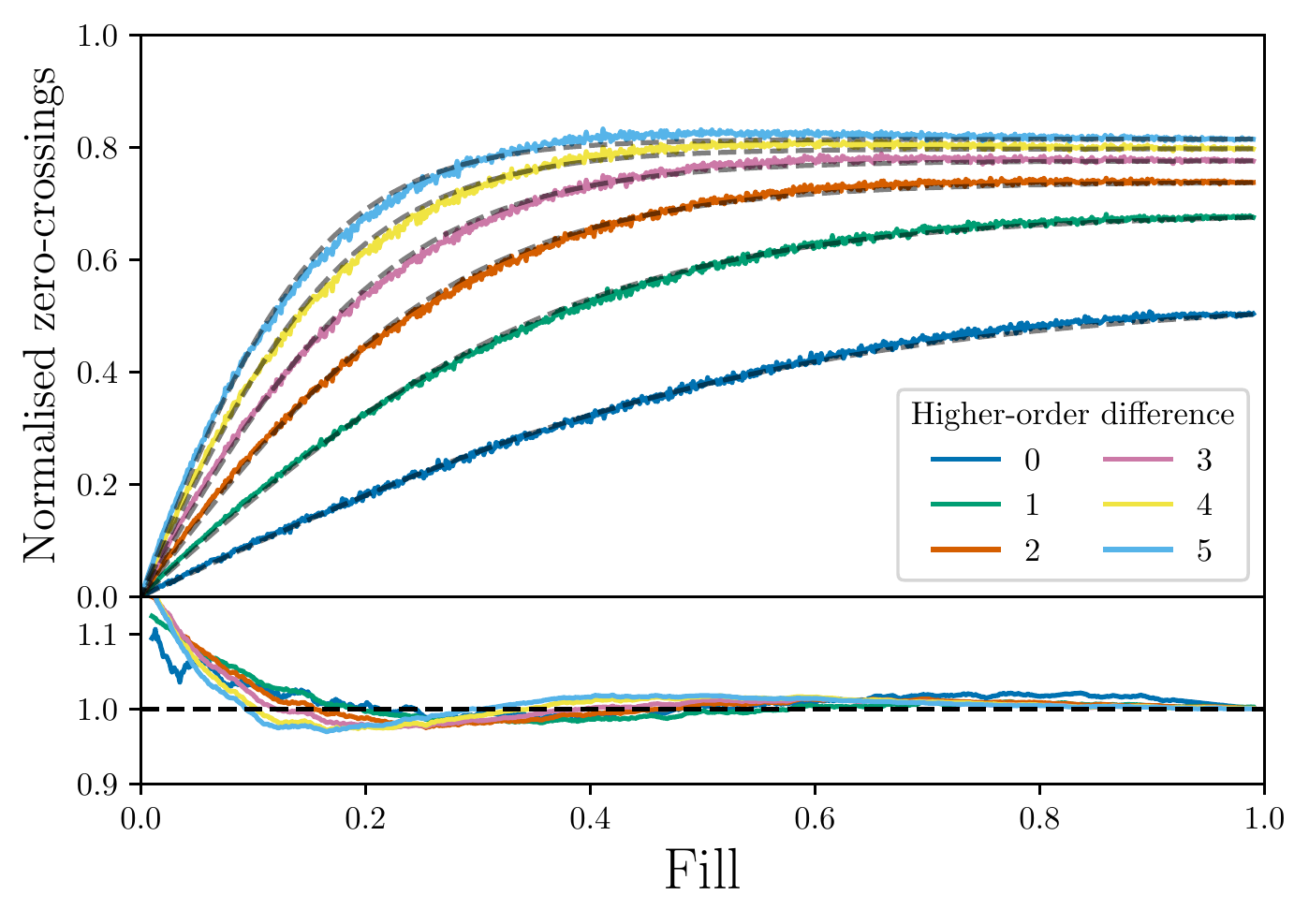}
\caption{The number of zero crossings in a white noise time-series normalised by the number of data points in the data, plotted as a function of fill. Each line is coloured according to the higher-order difference calculated, where 0 is the equivalent to the zero-crossings of the original time-series data. The normalisation functions for each higher-order difference are given by the grey dashed lines. The bottom panel shows the residuals of the normalisation function fits to the data in the top panel, smoothed over 20 points for clarity and coloured by the corresponding higher-order difference.}
\label{fig:hoc_fill}
\end{figure}

The correction factor was determined using a number of simulated white noise time-series. Fig.~\ref{fig:hoc_fill} shows the results of the simulations for a variety of fill values and higher-order differences. In order to produce data with differing numbers of gaps, data points were randomly selected and set to zero, in accordance with the desired fill value. As a result only short timescale gaps were inserted.

The shape of the curves in the top panel of Fig.~\ref{fig:hoc_fill} resemble half of a sigmoid function, which provides a straight-forward analytical formulation that can be fitted to the data. We define the normalisation function as follows
    
\begin{equation}
	S(x) = \frac{1}{1 + \exp(-kx)} - 0.5,  \quad \text{for } x > 0,
\end{equation}
where $x$ is the value of the fill and $k$ is a free parameter that is fitted to the data.

The subtraction of 0.5 ensures $S(x=0)=0$, which is a condition set by our data since when there are no data points (i.e. all zeros), there are no zero-crossings. However in its current guise, the above function is not quite suitable because it varies between 0 and 0.5 for positive $x$. In order to extend it to the full range over the number of normalised zero-crossings we normalise it by $S(x=1)$, the value of the function for full fill, and then multiply it by the number of normalised zero-crossings at full fill for each higher-order crossings (HOC). Examples of this function fitted to the data are shown in Fig.~\ref{fig:hoc_fill}, which was performed by minimising the residual sum of squares.

The fill correction is a multiplicative factor and so to discern the accuracy of the fitted values we look at the ratio of the data to the fill correction as a function of fill, as shown in the bottom panel of Fig.~\ref{fig:hoc_fill}. Note the scale of the ``residuals'' which in turn shows that we are precise to $\sim5$\% (for the worst case) at a fill of 0.4. Due to the additional data pre-processing and the merging of data with large gaps in between, the fill of the real data rarely falls below $\sim0.8$.

\section{Hyperparameter tuning}\label{sec:hyperparams}

There are a number of hyperparameters of the model that require tuning in order to achieve a high degree of accuracy.

\begin{itemize}
    \item the learning rate $\eta\in[0,1]$ dictates how much the contribution of the newest tree is scaled relative to the current model, i.e. how much of the variance of residuals it is fitted to. A lower value removes less variance from the residuals and indicates a more conservative model.
    \item \texttt{max\_depth} controls the maximum depth of each tree, i.e. how many times an individual tree undergoes a split.
    \item \texttt{min\_child\_weight} controls the conditions under which a split occurs. Only if the sum of the weights at a given split exceeds this value does the tree continue to grow.
    \item \texttt{subsample} dictates the fraction of training data used to grow each tree.
    \item \texttt{colsample} which dictates the fraction of features used when constructing each tree. This means that not all features are necessarily used when constructing each tree.
    \item $\lambda$ is the L2 regularisation term on the model weights and is fixed to the default value of 1.
    \item $\alpha$ is the L1 regularisation term on the model weights and is fixed to the default value of 0.
\end{itemize}

The degree of regularisation, $\gamma$, which penalises the complexity of each constructed tree is not included in the hyperparameter tuning. This is because a value of zero, i.e. no regularisation, will always be chosen as introducing regularisation will inherently increase the minimum loss value that can be obtained. The value of $\gamma$ was chosen such that the difference between the accuracy of the training and validation sets were minimised, i.e. the model generalises well to the unseen validation data. The value of $\gamma$ was fixed across all lengths of dataset to a value of 7.5.

We use the \texttt{hyperopt} package \citep{pmlr-v28-bergstra13} to perform the search over the hyperparameter space defined in Table~\ref{table:hyper}. The \texttt{hyperopt} package then finds the combination of hyperparameters that minimises a chosen objective function using a random search. This function is chosen to be the multi-class log-loss (as given in Eq. \ref{eqn:logloss}) and 250 trials are made to find the best combination of hyperparameters. The search is performed separately for each time-series length.

\begin{table*}
	\centering
	\caption{The \texttt{xgboost} hyperparameter values and hyperparameter space chosen for tuning.}
	\label{tab:hyperparams}
	\begin{tabular}{lcccccc} 
	\hline
	& \multicolumn{4}{c}{Length of dataset} & \multicolumn{2}{c}{Tuning}\\
	Hyperparameter & 4 years & 180 days & 80 days & 27 days & Range & Step\\
	\hline
	$\eta$ & 0.1 & 0.375 & 0.275 & 0.025 & 0.025-0.5 & 0.025\\
	$\gamma$ & 7.5 & 7.5 & 7.5 & 7.5 & - & - \\
	\texttt{max\_depth} & 3 & 6 & 10 & 8 & 1-13 & 1\\
	\texttt{min\_child\_weight} & 4 & 1 & 2 & 3 & 1-6 & 1\\
	\texttt{subsample} & 0.7 & 0.5 & 0.85 & 0.5 & 0.5-1.0 & 0.05\\
	\texttt{colsample} & 0.7 & 0.5 & 0.85 & 0.85 & 0.5-1.0 & 0.05\\
	\hline
	\end{tabular}
\label{table:hyper}
\end{table*} 

\section{Probability Calibration}\label{sec:prob_calib}

As stated previously the classifier returns the probability that a star belongs to a given class rather than just the class label. This gives the possibility for the probability produced by the classifier to be used in subsequent probabilistic analyses, whether that be as a prior distribution on RC membership or perhaps in future peak-bagging codes. In order for these probabilities to be used they need to be well-calibrated so that they can be used as prior or posterior probabilities.

The classifier produces a probability that a star belongs to a given class that we shall call the ``confidence'', if the classifier assigns a confidence of 0.5 to 100 predictions then we would expect 50 of the predictions to be correct. If this is the case then the probabilities are said to be ``calibrated''. 

It is important to assess whether the probabilities outputted by the classifier are well-calibrated. In this work we use temperature scaling applied to the multiclass classification problem from \cite{Guo:2017:CMN:3305381.3305518} \footnote{\url{https://github.com/gpleiss/temperature_scaling}} to test the calibration of the probabilities. The way in which the classifier obtains class probabilities is by predicting the logits (the logarithm of the odds that a star belongs to a particular class) and passing them through a softmax function
\begin{equation}
    \sigma_{\mathrm{SM}}(z_{i})^{(k)} = \frac{\exp(z_{i}^{(k)})}{\sum_{j=1}^{K}\exp(z_{i}^{(j)})},
\end{equation}
where $K$ is the number of classes in the classification problem, the index $i$ represents an individual star and $z_{i}$ are the logits for a given star. The class prediction probability is then made by
\begin{equation}
    \hat{p}_{i} = \underset{k}{\max}\;\sigma_{\mathrm{SM}}(z_{i})^{(k)}.
\end{equation}
The aim of temperature scaling is to essentially ``soften'' the softmax function changing the output probabilities, but not changing to the overall accuracy of the model. In other words we want to obtain new calibrated probabilities $\hat{q}$ such that
\begin{equation}
    \hat{q}_{i} = \underset{k}{\max}\;\sigma_{\mathrm{SM}}(z_{i}/T)^{(k)},
\end{equation}
where $T$ is the temperature that is a fitted parameter post classifier training (see \citealt{Guo:2017:CMN:3305381.3305518} for more details).

\begin{figure*}
\centering
\subfloat[\label{fig:pre_cal}]{\includegraphics[width=0.45\textwidth]{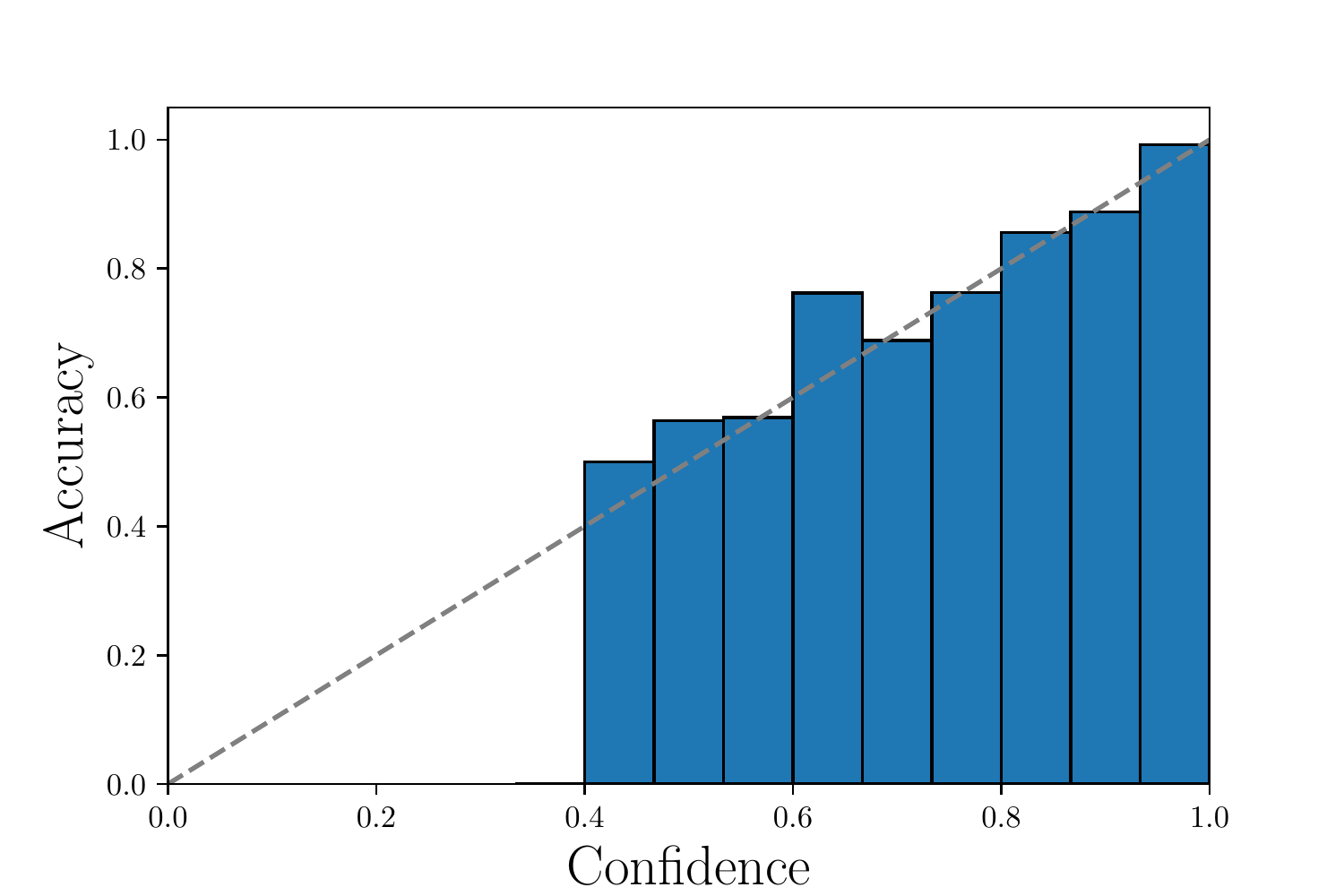}}
\subfloat[\label{fig:post_cal}]{\includegraphics[width=0.45\textwidth]{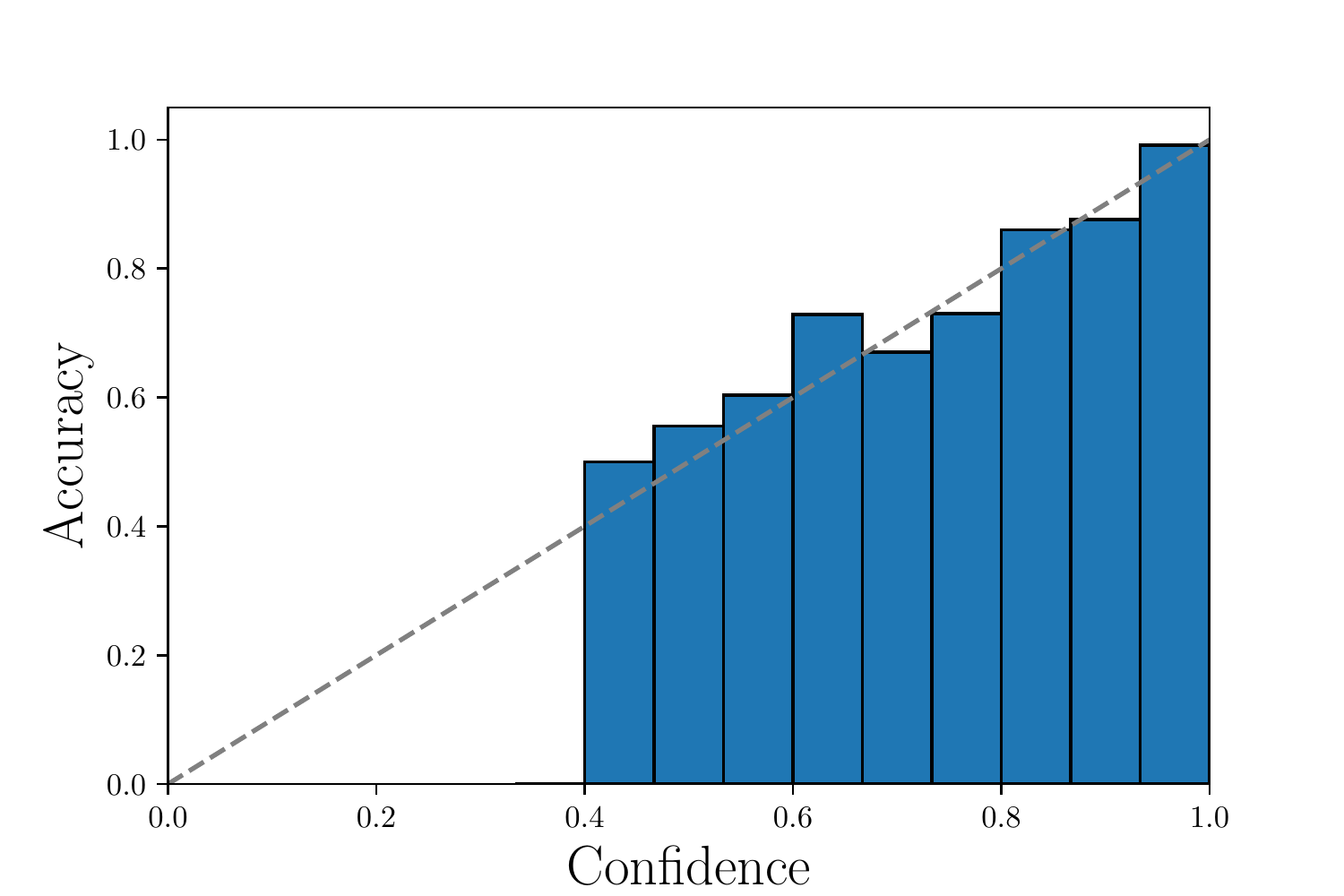}}
\caption{A reliability plot for the 27 day classifier shown before the probability calibration (a) and after the calibration (b). The 1:1 diagonal is given by the grey dashed line.} 
\label{fig:reliability}
\end{figure*}

The calibration can be summarised in a reliability plot, as shown in Fig~\ref{fig:reliability}, which displays the binned confidence of the predictions from a classifier as a function of the accuracy of the classifier in that bin. If the probabilities are perfectly calibrated then the data will lie along the 1:1 diagonal, and any deviation will show some degree of miscalibration. Fig.~\ref{fig:pre_cal} shows the reliability plot of the 27 day classifier before the probability calibration and it can be seen that the probabilities are well calibrated. Once the probability calibration has been applied the difference is only very slight. As a result we choose not to add this extra post-processing step to the classifier probabilities and confirm that the probabilities can be reliably interpreted in subsequent analyses.

\bsp	
\label{lastpage}
\end{document}